\documentclass[12pt]{article}

\usepackage[utf8]{inputenc}

\usepackage{amsmath}
\usepackage{amsfonts}
\usepackage{graphicx}
\usepackage{xcolor}
\usepackage{enumitem}
\usepackage[export]{adjustbox}
\usepackage{amssymb,latexsym,amsthm, mathrsfs}
\newtheorem{theorem}{Theorem}[section]

\theoremstyle{definition}

\makeatletter
\usepackage{authblk}
\usepackage{graphics}
\usepackage{epsfig}
\usepackage{tabularx,caption,subcaption,graphicx}
\usepackage{amsthm,dsfont,commath,tabulary}

\begin{document}


  \title{Comparison of the analytical approximation formula and Newton's method for solving a class of nonlinear Black-Scholes parabolic equations
  \thanks{The final publication is available at www.degruyter.com once it is published}
  }
  
  \author[1]{Karol \v{D}uri\v{s}}
  \author[2]{Shih-Hau Tan}
  \author[3]{Choi-Hong Lai} 
  \author[4]{Daniel \v{S}ev\v{c}ovi\v{c}}
  \affil[1]{\small National Bank of Slovakia, Slovakia, e-mail: {duris.karol@gmail.com}}
  \affil[2]{\small Department of Mathematical Sciences, University of Greenwich, London, UK, e-mail: {s.tan@gre.ac.uk}}  
  \affil[3]{\small Department of Mathematical Sciences, University of Greenwich, London, UK, e-mail: {C.H.Lai@greenwich.ac.uk}}
  \affil[4]{\small Department of Applied Mathematics and Statistics, Division of Applied Mathematics, Comenius University, Bratislava, Slovakia, e-mail: {sevcovic@fmph.uniba.sk}}
  
\date{}

  

\maketitle

  \abstract{Market illiquidity, feedback effects, presence of transaction costs, risk from unprotected portfolio and other nonlinear effects in PDE based option pricing models can be described by solutions to the generalized Black-Scholes parabolic equation with a diffusion term nonlinearly depending on the option price itself. Different linearization techniques such as Newton's method and analytic asymptotic approximation formula are adopted and compared for a wide class of nonlinear Black-Scholes equations including, in particular, the market illiquidity model and the risk-adjusted pricing model. Accuracy and time complexity of both numerical methods are compared. Furthermore, market quotes data was used to calibrate model parameters.


 
  }
  
\section{Introduction}
According to the classical theory due to Black, Scholes and Merton an option in a stylized and idealized financial market can be priced by a solution $V=V(S,t)$ to the linear Black--Scholes parabolic equation:
\begin{equation}
\frac{\partial V}{\partial t}  + \frac{1}{2}\tilde\sigma^2 S^2 \frac{\partial^2 V}{\partial S^2} +  (r-q) S\frac{\partial V}{\partial S} - r V =0,
\label{doplnky-c-BS}
\end{equation}
where $r>0$ is the interest rate of a zero-coupon bond, $q\ge 0$  is the dividend yield rate and $\tilde\sigma>0$ is a constant historical volatility of the underlying asset price process $\{S_t, t\ge0\}$, which is assumed to follow a stochastic differential equation
\begin{equation}
d S_t = (r-q) S_t dt + \tilde\sigma S_t d W_t,
\label{doplnky-afm-geomBrown}
\end{equation}
of the geometric Brownian motion with a drift $r-q$ (cf. \cite{Kw,H,Daniel1}). The linear Black--Scholes equation with a constant volatility $\tilde\sigma$ has been derived under several restrictive assumptions, for example, zero transaction costs, perfectly replicated portfolio, frictionless, market completeness, etc.

In this paper, the main goal is to investigate and compare two numerical approximation methods for solving a class of nonlinear generalizations of the linear Black-Scholes equation (\ref{doplnky-c-BS}) in which the volatility is assumed to be a function of the underlying asset price $S$ and Gamma of the option (the Greek Gamma is the second derivative $\partial^2_S V$), i.e. 
\begin{equation}
\sigma = \sigma(\partial^2_S V, S)\,.
\label{doplnky-c-sigma}
\end{equation}
The motivation for solving the nonlinear Black--Scholes equation (\ref{doplnky-c-BS}) with the volatility function $\sigma$ of the form (\ref{doplnky-c-sigma}) arises from more realistic option pricing models in which one can take into account nontrivial transaction costs, market feedbacks, risk from unprotected portfolio and other effects. In the last decades, some of the restrictive assumption of the classical Black--Scholes theory \cite{BS} have been relaxed in order to model, for instance,  presence of constant transaction costs (see e.g. Leland \cite{doplnkyLe}, Hoggard {\em et al.} \cite{HWW}), non-constant transaction costs (see e.g. Amster  {\em et al.} \cite{AAMR}, \v{S}ev\v{c}ovi\v{c} and \v{Z}it\v{n}ansk\'a \cite{SZ}), uncertain volatility model (cf. Avellaneda and  Paras \cite{doplnkyAP}), feedback and illiquid market effects due to large traders choosing given stock-trading  strategies (cf. Frey \cite{Frey}, Frey and Patie \cite{FP}, Frey and Stremme \cite{FS}, Sch\"onbucher and  Wilmott \cite{SW}), imperfect replication and investor's preferences (cf. Barles and Soner  \cite{BaSo}), risk from an unprotected portfolio (cf. Kratka \cite{doplnkyKr}, Janda\v{c}ka and \v{S}ev\v{c}ovi\v{c} \cite{Daniel2}). Efficient techniques and fast computational methods for pricing derivative securities is a practical task in financial quotes markets. Therefore, realistic PDE based option models including, in particular, nonlinear generalizations of the Black--Scholes equation have to be solved in a fast and efficient way. However, in most important cases there is no explicit formula except for some special cases with non-standard pay-off diagrams (cf. Bordag \cite{BordagBook}). This is the reason why numerical methods for solving nonlinear Black--Scholes equation have to be developed and analyzed.

In this paper, attention is focused on a class of nonlinear Black-Scholes equations. In particular, the nonlinear volatility model developed by Frey \emph{et al.} \cite{Frey,Frey1998,FP,FS} and the risk-adjusted pricing methodology model proposed and investigated by Kratka \cite{doplnkyKr} and Janda\v{c}ka and \v{S}ev\v{c}ovi\v{c} \cite{Daniel2,Daniel1} are the main concern of this work. In a series of papers \cite{Frey,Frey1998,FP,FS} Frey  \emph{et al.} considered a model in which the price of an underlying asset is affected by specific hedging strategies due to a large trader. Supposing that a large trader uses a given stock-holding strategy $\alpha_t$ and the underlying stock price process satisfies the following SDE:
\begin{equation}
dS_t = \mu S_t dt + \sigma S_t d W_t + \rho S_t d\alpha_t,
\label{frey-S}
\end{equation}
where $\mu$ is a drift parameter, $\sigma>0$ is the volatility of the process and $0\le \rho <\bar\rho$ is the so-called market liquidity parameter. It is worth noting that the quantity $1/(\rho S_t)$ measures the size of the change in the stock-holding position of the large trader. Notice that if $\alpha_t\equiv 0$ or $\rho=0$, the stock price $S_t$ follows the geometric Brownian motion. In \cite{Frey} Frey (see also \cite{FP,FS}) showed that the option price is then a solution to a nonlinear volatility Black-Scholes equation of the form:
\begin{equation} \label{nonBS}
\frac{\partial V}{\partial t} + \frac{1}{2}\sigma(\partial^2_S V, S)^2 S^2\frac{\partial^2 V}{\partial S^2} + (r-q)S\frac{\partial V}{\partial S}  -rV = 0,
\end{equation}
for $0 \leq S < \infty$ and $ 0  \leq t < T$ where $T$ is the maturity time. The nonlinear volatility function $\sigma$ is given by
\begin{equation} \label{Nonlinearity}
\sigma(\partial^2_S V, S) = \tilde{\sigma}(1-\rho S\partial^2_S V)^{-1},
\end{equation}
where $\tilde{\sigma}$ is a constant historical volatility.  A solution $V=V(S,t)$ is subject to the terminal pay-off condition describing call or put option with expiration price $E>0$, i.e. 
\begin{equation}
V(S,T) = (S-E)^+ \quad\hbox{(call option),}\quad
V(S,T) = (E-S)^+ \quad\hbox{(put option).}
\label{payoff}
\end{equation}

Another nonlinear model was proposed by Kratka \cite{doplnkyKr}. It was further generalized and analyzed by Janda\v{c}ka and \v{S}ev\v{c}ovi\v{c} in \cite{Daniel2,Daniel1}. The model is constructed following the classical Leland approach for modeling transaction costs (cf. \cite{doplnkyLe}) in which the time between consecutive portfolio rearrangements is subject to optimization with respect to the risk arising from an unprotected portfolio. In this risk-adjusted pricing methodology (RAPM) model the nonlinear volatility function  has the form:
\begin{equation} \label{Nonlinearity_RAPM}
\sigma(\partial^2_S V, S)^2 = \tilde{\sigma}^2(1+\mu (S\partial^2_S V)^{\frac{1}{3}}).
\end{equation}

Construction of explicit solutions to equation (\ref{nonBS}) with the nonlinear volatility function as the one defined in (\ref{Nonlinearity_RAPM}) were recently provided by Bordag and Frey \cite{Bordag} (see also \cite{BordagBook}). Several invariant solutions were constructed by means of the invariant Lie group theory. These invariant solutions depend on various parameters restricting the class of solutions. In particular, not every pay-off diagram can be considered. In general, there is no exact pricing formula for the case of a call or put terminal pay-offs. Hence efficient numerical techniques for solving such nonlinear Black--Scholes equations are required.

A numerical method proposed and investigated by Janda\v{c}ka and \v{S}ev\v{c}ovi\v{c} \cite{Daniel2} is based on the transformation $H=S\partial^2_S V, x=\ln(S/E), \tau=T-t$, which transforms equation (\ref{nonBS}) with $\sigma=\sigma(S\partial^2_S V)$ into a porous media type of quasilinear parabolic equation:
\begin{equation}
\frac{\partial H}{\partial \tau}
= \frac{\partial^2}{\partial x^2} \beta(H) + \frac{\partial}{\partial x}\beta(H)
+ (r-q) \frac{\partial H}{\partial x} - q H\,,
\label{doplnky-jam-equationH}
\end{equation}
where $\beta(H) = \frac{1}{2}\sigma^2(H)H$ is an increasing function. For instance, in the case of the volatility function given by (\ref{Nonlinearity})  one obtains $\beta(H)=\frac{\tilde\sigma^2}{2} H \left(1- \rho H\right)^{-2}$ for $H<H_{max}$ (see \cite{Daniel1} for details). In the recent paper \cite{SZ}, \v{S}ev\v{c}ovi\v{c} and \v{Z}it\v{n}ansk\'a investigated the nonlinear equation (\ref{doplnky-jam-equationH}) in the context of modeling variable transaction costs. The existence of classical H\"older smooth solutions was proved and useful bounds for the solution were derived. 

The transformation technique developed in \cite{Daniel2} allows for construction of a semi-implicit finite volume based numerical scheme for solving  (\ref{doplnky-jam-equationH}). There are other approaches dealing mainly with the nonlinear equation (\ref{nonBS}) for the option price rather than for its transformation $H=S\partial^2_S V$. Another method using quasilinearization technique for solving the fully nonlinear parabolic equation (\ref{nonBS}) was proposed and analyzed by Koleva and Vulkov \cite{Koleva}. A consistent monotone explicit finite difference numerical scheme was analyzed by Company \emph{et al.} in the context of the Frey and Patie model (\ref{nonBS}) with a nonlinear volatility function given by (\ref{Nonlinearity}). In \cite{Matthias2} Ehrhardt and Valkov  derived an unconditionally stable explicit numerical scheme for solving the same problem and provided necessary numerical analysis of the scheme. 

In this paper, two numerical approximation methods based on the asymptotic perturbation analysis and the Newton linearization technique are developed. These methods are used to solve a wide class of nonlinear Black-Scholes equations. The first method is the asymptotic perturbation method which is based on asymptotic expansion of the solution into power series in a small model parameter. The first order expansion then corresponds to an explicit analytic approximation formula requiring only one-dimensional numerical integration which can be computed in a fast and efficient way. The second method is based on Newton's iterative method for solving the corresponding nonlinear problem in each temporal discretization level. It is applicable to a rather general nonlinear case not restricted by any specific types of equations and boundary and terminal conditions.  In \cite{Heider}  Heider used Newton's iterative method for solving equation (\ref{nonBS}) with four types of nonlinear volatilities and different finite different schemes.  Note that different variants of Newton's linearization and their implementation are also discussed and compared in this paper.

The paper is organized as follows. In section 2 an explicit analytic approximation formula for solving a general class of nonlinear volatility models is derived. In section 3 an algorithm utilizing Newton's method for solving equation (\ref{nonBS}) is described and analyzed. Several comparisons of both methods are discussed in section 4. Examples of solution to the Frey and Patie model and RAPM model are presented. Finally, section 5 contains an example of model calibration to real market quotes data.

\section{Analytic approximation formula based asymptotic perturbation analysis}
In this section, an analytic approximation formula for pricing European call or put options with a nonlinear volatility is derived. Typically this paper considers a wide class of nonlinear volatility functions taking the following form:
\begin{equation*}\label{vol}
\sigma(\partial^2_S V,S,T-t)^2= \tilde{\sigma}^2 + 2 \varepsilon A(T-t) S^{\gamma-1} H^{\delta-1}, \quad \hbox{where}\quad H=S\frac{\partial^2 V}{\partial S^2}.
\end{equation*}
The powers $\gamma,\delta$, the parameter $\varepsilon$ as well as the function $A(T-t)$ depend on the chosen nonlinear volatility model.  For example, in the case of the Frey and Patie model with the nonlinear volatility function given by (\ref{Nonlinearity}), $\sigma(\partial^2_S V, S) = \tilde \sigma/(1-\rho S \partial^2_S V) \approx \tilde \sigma (1 +\rho S \partial^2_S V)$ 
and the parameters are 
\[
\varepsilon=\rho, \gamma=1, \delta=2, A(T-t)=\tilde\sigma^2,
\]
and the small model parameter $\varepsilon$ can be identified with $0<\rho\ll 1$.

For the RAPM model with the nonlinear volatility function given by (\ref{Nonlinearity_RAPM}) the parameters can be identified as follows:
\[
\varepsilon=\mu, \gamma=1, \delta=4/3, A(T-t) = \tilde\sigma^2/2, 
\]
and the small parameter $\varepsilon$ is identified with $0<\mu\ll 1$.

Equation (\ref{nonBS}) can now be rewritten as
\begin{align}\label{BSzovseobMMP}
  \begin{split}
    &\mathcal{L}(V,\varepsilon)\equiv \frac{\partial V}{\partial t} + \frac{1}{2}\sigma(\partial^2_S V,S)^2 S^2\frac{\partial^2 V}{\partial S^2} + (r-q)S\frac{\partial V}{\partial S}  -rV = 0,  \\
    &V(S,T)=\bar V(S),
  \end{split}
\end{align}
where $\bar V$ is the prescribed pay-off diagram. The problem is to seek the option price in the form of an asymptotic expansion in terms of a small parameter (cf. \cite{holmes2}). More precisely, 
\begin{equation}\label{Rozvoj}
V = V_0 + \sum\limits_{i=1}^{N} \varepsilon^{i} V_i + O(\varepsilon^{N+1}),
\end{equation}
where the leading term $V_0\equiv V_{BS}$ is simply a solution to the linear Black-Scholes model. 

The aim here is to derive an asymptotic approximation formula obtained from the first two terms in the asymptotic expansion, i.e. 
\begin{equation}\label{RozvojBS}
  V(S,t)\approx V_0(S,t)+\varepsilon V_1(S,t).
\end{equation}
In order to obtain an explicit formula for the second term $V_1$ in the expansion, equation (\ref{BSzovseobMMP}) is first approximated as follows:
\begin{equation}\label{ExponMMPskrat}
\mathcal{L}(V,\varepsilon)
\approx\mathcal{L}_0(V)+\varepsilon\mathcal{L}_1(V)
\approx\mathcal{L}_0(V_0+\varepsilon V_1)+\varepsilon\mathcal{L}_1\left(V_0+\varepsilon V_1\right),
\end{equation}
where $\mathcal{L}_0$ is a linear and $\mathcal{L}_1$ is a nonlinear differential operator in $V$,
\begin{eqnarray*}
&&\mathcal{L}_0(V)\equiv \frac{\partial V}{\partial t}+\frac{1}{2} \tilde{\sigma}^2 S^2 \frac{\partial^2 V}{\partial S^2}+(r-q)S\frac{\partial V}{\partial S} -rV,
\\
&&\mathcal{L}_1(V)\equiv A(T-t) S^\gamma \left(S\frac{\partial^2 V}{\partial S^2}\right)^\delta.
\end{eqnarray*}

Hence the first order approximation of the equation $\mathcal{L}(V,\varepsilon)=0$ reads as follows:
\begin{equation}\label{BSmmpOdv3}
\mathcal{L}_0(V_0)+\varepsilon(\mathcal{L}_0(V_1)+\mathcal{L}_1(V_0)) =0
\end{equation}
satisfying the initial condition: 
\begin{equation}\label{BSmmpPoc}
  V(S,T)\equiv V_0(S,T)+\varepsilon V_1(S,T)= \bar V(S). 
\end{equation}
Equation (\ref{BSmmpOdv3}) with the initial condition (\ref{BSmmpPoc}) can be separated into a system of equations in powers of $\varepsilon$, i.e.
\begin{eqnarray*}
O(\varepsilon^0): &&\mathcal{L}_0(V_0)=0,
\\
&& V_0(S,T)=(S-E)^+\quad \hbox{(call)},\quad V_0(S,T)=(E-S)^+ \quad \hbox{(put)}
\\
\\
O(\varepsilon): &&\mathcal{L}_0(V_1)=-\mathcal{L}_1(V_0), 
\\
&& V_1(S,T)=0.
\end{eqnarray*}
The solution $V_0$ can be obtained by solving the linear Black-Scholes equation. The second equation for $V_1$ is a non-homogeneous PDE with zero initial condition. 

Introduce $H_0=S\frac{\partial^2 V_0}{\partial S^2}$, the equation $\mathcal{L}_0(V_1)=-\mathcal{L}_1(V_0)$ can be rewritten as follows
\begin{equation}\label{zadanie}
 \left\{
  \begin{array}{ll}
    \mathcal{L}_0(V_1)=-A(T-t)S^\gamma H_0^\delta, & (S,t)\in(0,\infty)\times [0,T), \\
    V_1(S,T)=0, & S\in(0,\infty).
  \end{array}
  \right.
\end{equation}

Therefore, equation (\ref{zadanie}) can be solved once the value of $V_0(S,t)$ is evaluated to obtain $H_0$.  Recall
\begin{eqnarray*}
&&V_0(S,t)= Se^{-q(T-t)}\Phi(d_1)-Ee^{-r(T-t)}\Phi(d_2), \\
&&d_{1,2}=\frac{\ln\frac{S}{E}+\left(r-q\pm\frac{\tilde{\sigma}^2}{2}\right)(T-t)}{\tilde{\sigma}\sqrt{T-t}},
\end{eqnarray*}
where $\Phi(d)=\frac{1}{\sqrt{2\pi}}\int_{-\infty}^d e^{-x^2/2}dx$ is the cumulative distribution function of the standard normal distribution. Hence
\begin{equation*}
H_0 = S \frac{\partial^2V_0}{\partial S^2}=\frac{e^{-q\tau}\Phi'(d_1)}{\tilde{\sigma}\sqrt{\tau}}.
\end{equation*}

In order to solve equation (\ref{zadanie}) one adopts the usual transformation (see e.g. \cite{Daniel1})
\begin{equation}\label{transf}
  \tau=T-t,\quad  S=Ee^x, \quad  e^{\alpha x+\beta\tau}u(x,\tau)=V_1(S,t),
\end{equation}
where 
\begin{equation}\label{alfa_beta}
   \alpha=\frac{1}{2}+\frac{q-r}{\tilde{\sigma}^2}, \quad
    \beta=-\left(\frac{\tilde{\sigma}^2}{8}+\frac{r+q}{2}+\frac{(r-q)^2}{2\tilde{\sigma}^2}\right)= -\frac{\tilde{\sigma}^2}{2}\alpha^2-r.
\end{equation}
Equation (\ref{zadanie}) is thus transformed to as follows:
\begin{align}\label{transfzadanie}
  \begin{split}
    -e^{\alpha x+\beta\tau}&\frac{\partial u}{\partial \tau} +e^{\alpha x+\beta\tau}\frac{\tilde{\sigma}^2}{2} \frac{\partial^2 u}{\partial x^2}= -A(\tau)\,\,\,E^\gamma e^{\gamma x}\,\,\, e^{-q\delta\tau}\frac{\left(\Phi'(\tilde{d_1})\right)^\delta}{\tilde{\sigma}^\delta \tau^{\frac{\delta}{2}}}, \\
    &u(x,0)=0.
  \end{split}
\end{align}
The term $\tilde{d_1}$ corresponds to $d_1$ after transformation (\ref{transf}). It is given by 
\[
  \tilde{d_1}=\frac{x}{\tilde{\sigma}\sqrt{\tau}}+\frac{\left(r-q+\frac{\tilde{\sigma}^2}{2}\right)}{\tilde{\sigma}}\sqrt{\tau}= \frac{x}{\tilde{\sigma}\sqrt{\tau}}+(1-\alpha)\tilde{\sigma}\sqrt{\tau}.
\]
Finally, equation (\ref{transfzadanie}) can be simplified to as the equation below
\begin{eqnarray}\label{zakluloha}
 && \frac{\partial u}{\partial \tau} -\frac{\tilde{\sigma}^2}{2}\frac{\partial^2 u}{\partial x^2} = \frac{E^\gamma A(\tau)}{\left(2\pi\tilde{\sigma}^2\tau\right)^ {\frac{\delta}{2}}} e^{-\frac{\delta}{2\tilde{\sigma}^2\tau}x^2+\left[\gamma-\delta-\alpha(1-\delta)\right]x 
 -\left[\beta+q\delta+\frac{\delta}{2}(1-\alpha)^2\tilde{\sigma}^2\right]\tau } \nonumber,
\\
&& u(x,0)=0, \quad (x,\tau)\in\mathds{R}\times[0,T].
\end{eqnarray}

\begin{theorem}\label{maintheorem}
Let $u(x,\tau)$ be a solution to (\ref{zakluloha}) satisfying the growth condition $|u(x,\tau)|\leq M e^{b |x|^2}$ for all $x\in \mathds{R}, \tau\in[0,T]$ where $M,b$ are some constants. Then $u(x,\tau)$ is given by the formula:
\begin{equation}\label{Theorem}
u(x,\tau)=E^\gamma \int_0^\tau \frac{A(\xi)}{\Lambda(\tau,\xi)}e^{\left[\frac{P^2\tilde{\sigma}^2}{2(\delta-1)}+\beta(\delta-1)\right]\xi+\frac{Px}{1-\delta}+ \frac{P^2\tilde{\sigma}^2\tau}{2(1-\delta)^2}-\left[\frac{\delta x^2}{2\tilde{\sigma}^2}+ \frac{P x\delta\tau}{1-\delta}+ \frac{P^2\tilde{\sigma}^2\delta\tau^2}{2(1-\delta)^2}\right] \frac{1}{Q(\tau,\xi)}}d\xi,
\end{equation}
where $P=\gamma-\delta-\alpha(1-\delta)$ is a constant depending on the model parameters and the functions $Q(\tau,\xi)$ and $\Lambda(\tau,\xi)$ are defined as follows:
\begin{equation}\label{simplify}
Q(\tau,\xi)=\delta\tau+(1-\delta)\xi, \qquad \Lambda(\tau,\xi)=(2\pi\tilde{\sigma}^2)^{\frac{\delta}{2}}\xi^{\frac{\delta-1}{2}} \sqrt{Q(\tau,\xi)}.
\end{equation}
\end{theorem}

The proof of this theorem is a straightforward application of the variation of constants formula and can be found in the Appendix. As a consequence of the previous theorem an explicit expression for the first order approximation of the option price can be obtained. Taking $V_1(S,t)=e^{\alpha x+\beta\tau}u(x,\tau)$ leads to
\begin{eqnarray}\label{vypries5}
V_1(S,t)&=&\frac{E^\gamma}{(2\pi\tilde{\sigma}^2)^{\frac{\delta}{2}}}\left(\frac{S}{E}\right)^{\frac{\gamma-\delta}{1-\delta}}
    e^{\left\{\beta+\frac{\left[\gamma-\delta-\alpha(1-\delta)\right]^2\tilde{\sigma}^2}{2(1-\delta)^2}\right\}(T-t)}
    \\
    &&\times\int_0^{T-t} \frac{A(\xi)}{\xi^{\frac{\delta-1}{2}} \sqrt{\delta(T-t)+(1-\delta)\xi}} e^{K\xi-M(S)\frac{1}{\delta(T-t)+(1-\delta)\xi}}d\xi, \nonumber
\end{eqnarray}
where $K$ is a constant given by
\begin{equation*}\label{vyrazK}
  K=\frac{\left[\gamma-\delta-\alpha(1-\delta)\right]^2\tilde{\sigma}^2}{2(\delta-1)}+\beta(\delta-1)
\end{equation*}
and
\begin{eqnarray*}
M(S) &=&\frac{\delta}{2\tilde{\sigma}^2}\left(\ln\frac{S}{E}\right)^2+ \frac{\left[\gamma-\delta-\alpha(1-\delta)\right]\delta(T-t)}{1-\delta}\ln\frac{S}{E} \nonumber\\
  &&+ \frac{\left[\gamma-\delta-\alpha(1-\delta)\right]^2\tilde{\sigma}^2\delta(T-t)^2}{2(1-\delta)^2}.
\end{eqnarray*}
The analytic approximation of the option price $V(S,t)$ can then be evaluated by using equation (\ref{RozvojBS}).

\section{Implicit finite difference scheme using Newton's method}
A standard way of solving equation (\ref{nonBS}) numerically is to use implicit temporal discretization in combination with a finite difference method for approximating the derivatives. Note the volatility term appearing in (\ref{Nonlinearity}) and (\ref{Nonlinearity_RAPM}) is nonlinear and at each time level an iterative technique is to be applied. The method of frozen coefficient technique is commonly applied to handle the nonlinearity though sometimes it converges slowly without proper initial guess. To obtain a better convergence rate, Newton's method has to be employed in combination with a temporal implicit discretization scheme.

Newton's method is a linearization technique with many variants and each takes different implementation. In this section two approaches are discussed. The first approach (denoted by NM1) addresses the root-finding problem of the nonlinear system derived from an implicit scheme in which calculation of the Jacobian matrix is used to update the approximate solution. The second approach (NM2) linearizes the original equation in which a correction term is to be solved and used to update the approximate solution.

\subsection{Newton's Method (NM1)}
Using standard finite difference notations and the transformation $\tau = T - t$ an implicit finite difference scheme which replaces equation (\ref{nonBS}) reads as follows:
\[
\frac{V_i^{n+1} - V_i^n}{\Delta \tau} - \frac{1}{2} \sigma_i^{n+1}S_i^2 \frac{V_{i+1}^{n+1} - 2V_i^{n+1} +V_{i-1}^{n+1}}{(\Delta S) ^2} - rS_i \frac{V_{i+1}^{n+1} - V_{i-1}^{n+1}}{2\Delta S} + rV_i^{n+1} = 0.
\]
The volatility function $\sigma$ as given by (\ref{Nonlinearity}) may be discretised as
\[
\sigma_i^{n+1} = \tilde{\sigma}(1-\rho S_i \frac{V_{i+1}^{n+1}-2V_i^{n+1}+V_{i-1}^{n+1}}{(\Delta S)^2})^{-1}.
\]
Here $S_i = (i-1)\Delta S, \: i = 1,\cdots,M$, and $n = 1,\cdots,N-1$, where $M$ and $N$ are the numbers of grid points for spatial and temporal discretization respectively.  The above equation can be simplified as follows:
\begin{equation}
H(V^{n+1})V^{n+1} - V^n   = 0,
\end{equation}
where $H(V^{n+1})$ is an $M\times M$ tridiagonal matrix whose elements nonlinearly depend on $V^{n+1}$. Introducing the mapping 
\begin{equation}
G(V^{n+1}) = H(V^{n+1})V^{n+1} - V^n,
\end{equation}
turns the original problem to the construction of a solution $V^{n+1}$ of the equation $G(V^{n+1})=0$ at each time level.  Newton's method is applied to solve the root-finding problem which requires the Jacobian matrix of the function $G$ to be computed. An initial guess chosen as the solution $V$ from the previous time level usually reduces the number of Newton's iterations.
\\
\begin{center}
 \includegraphics{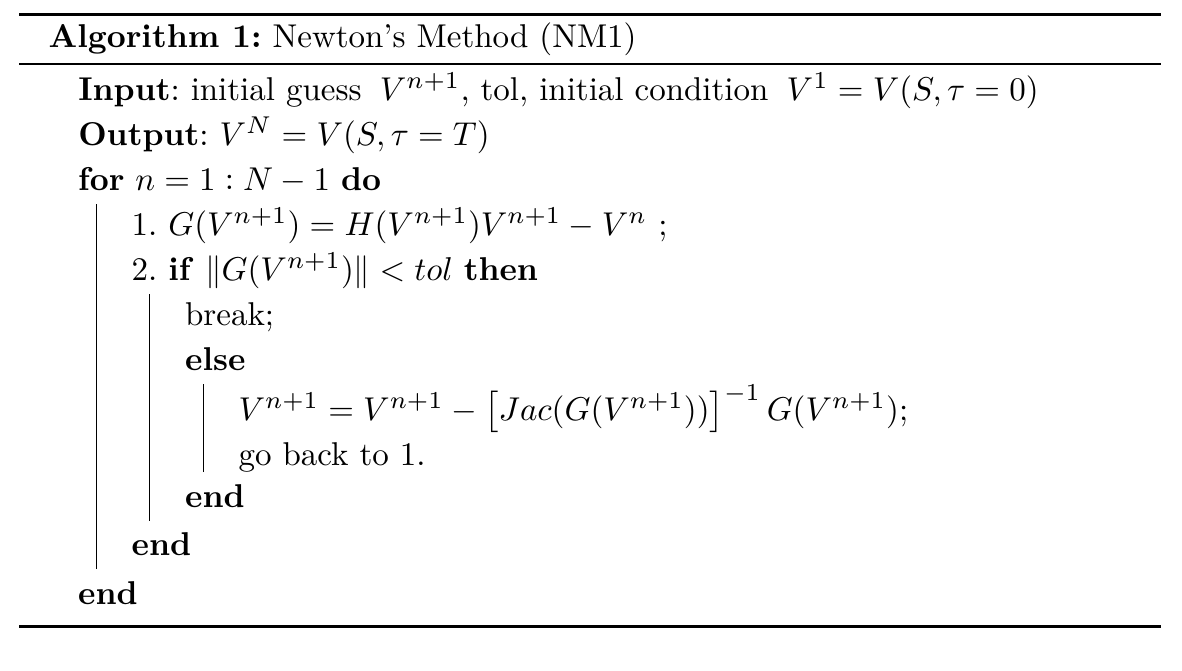}
\end{center}

In order to compute the Jacobian matrix efficiently, a decomposition of the matrix $H(V^{n+1})$ may be adopted as below 
\[
H(V^{n+1}) = \Sigma^{n+1}\,H_1 + H_2, \;\;\mbox{\rm where}\;\;  \Sigma^{n+1}= \mbox{\rm Diag}((\sigma^{n+1}_i)^2).
\]
Note that $H_1$ and $H_2$ are constant tridiagonal matrices.  By using this decomposition, the Jacobian matrix of $G$ becomes \\
\[
Jac(G(V^{n+1})) = 
\frac{\partial [H(V^{n+1})V^{n+1}]}{\partial V^{n+1}} = H(V^{n+1}) + \mbox{\rm Diag}( H_1\,V^{n+1} ) \nabla (\Sigma^{n+1})\, ,
\]
where
\[
\nabla (\Sigma^{n+1}) = ( (\nabla (\sigma_1^{n+1})^2)^T, (\nabla (\sigma_2^{n+1})^2)^T,...,(\nabla (\sigma_M^{n+1})^2)^T)^T\, ,
\]
here  $\nabla (\sigma_k^{n+1})^2$ is treated as a row vector.  

This decomposition simplifies the computing of the Jacobian matrix in terms of the nonlinear volatility. Each gradient $\nabla (\sigma_k^{n+1})^2$ can be obtained by either deriving the analytic expression for $\sigma$ or by using a finite difference approximation of the spatial derivatives. \\

\subsection{Waveform-Newton's Method (NM2)}
The second approach of applying Newton's linearization is to consider a smooth function $F$ representing the nonlinear Black-Scholes equation, i.e.
\[
F(V_{\tau},V_S,V_{SS},V) \equiv V_{\tau} - \frac{1}{2}\sigma^2(V_{SS}, S) S^2V_{SS}-rSV_S+rV =0.
\]
Here $V_{\tau},V_S,V_{SS}$ abbreviate the partial derivatives $\partial_\tau,\partial_S V, \partial^2_S V$ respectively. The linearization of the function $F$ at  $(V_{\tau}^*, V_S^*, V_{SS}^*,V^*)$ in direction  $(v_{\tau}, v_S, v_{SS},v)$ reads as follows:
\begin{equation} \label{NM2_eq}
\begin{split}
&F(V_{\tau}^*+v_{\tau},V_S^*+v_S,V_{SS}^*+v_{SS},V+v) \\&
= F(V_{\tau}^*,V_S^*,V_{SS}^*,V^*) + \frac{\partial F}{\partial V_{\tau}}v_{\tau} + \frac{\partial F}{\partial V_S}v_S + \frac{\partial F}{\partial V_{SS}}v_{SS} + \frac{\partial F}{\partial V}v+ O(D^2),
\end{split}
\end{equation}\\
where $D^2$  represents all higher order terms and the partial derivatives are evaluated at $(V_{\tau}^*,V_S^*,V_{SS}^*,V^*)$.  \\

Equation (\ref{NM2_eq}) transforms equation (\ref{nonBS}) into a linear partial differential equation of the correction term $v$ with zero boundary and initial conditions.  This equation can be solved easily because all coefficients of equation (\ref{NM2_eq}) are determined.  Similar to the first approach (NM1), these coefficients can be evaluated either by the analytic expression for $\sigma$ or by a finite difference approximation.  Eventually, the problem becomes
\begin{equation}\label{NM2_eq2}
\frac{\partial F}{\partial V_{\tau}^*} v^n - \Delta \tau F(V_{\tau}^*,V_S^*,V_{SS}^*,V^*) = H^*(V^*)v^{n+1}.
\end{equation}
Again, an initial guess can be set to the solution from previous time level in the algorithm.\\

\begin{center}
 \includegraphics{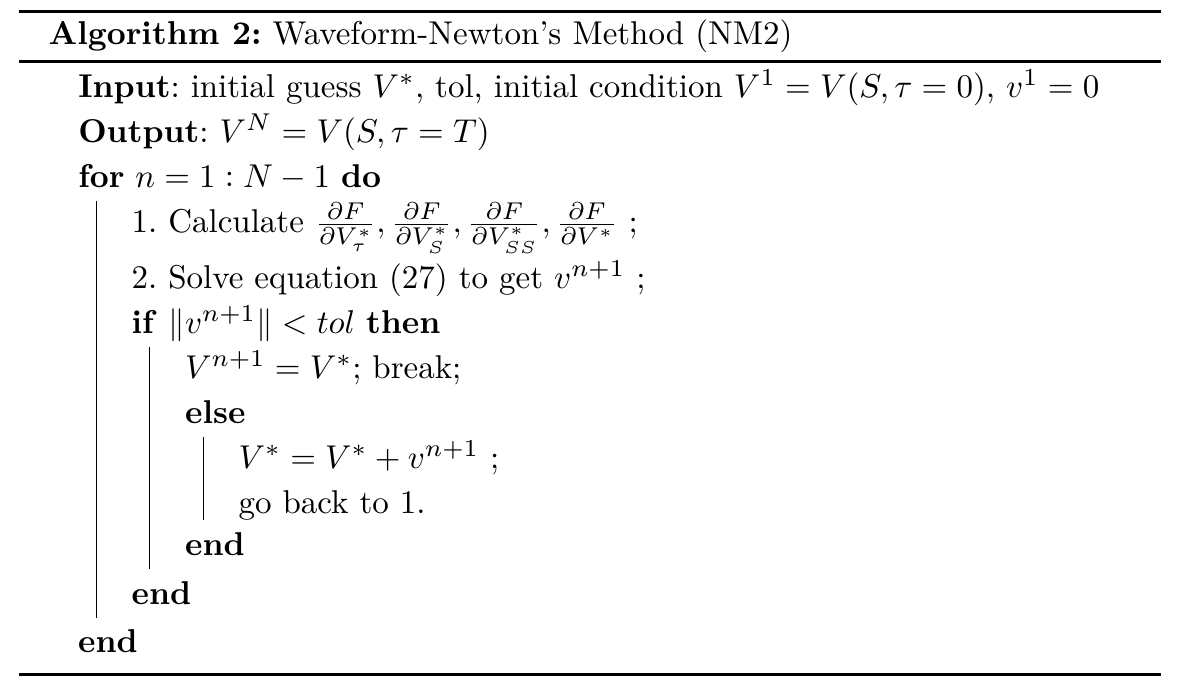}
\end{center}

\begin{figure} 
\begin{center}
\includegraphics[width=0.48\textwidth]{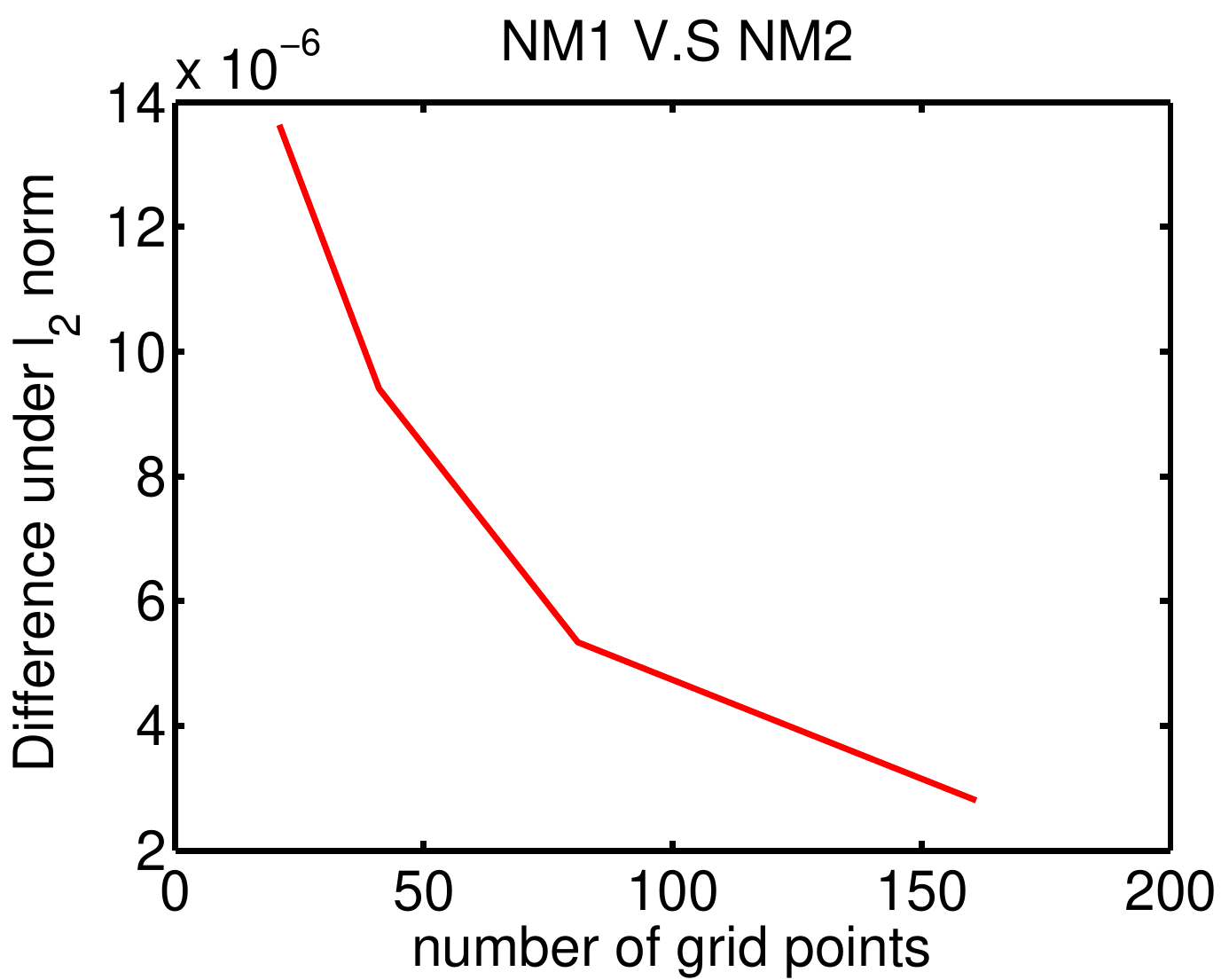}
\includegraphics[width=0.48\textwidth]{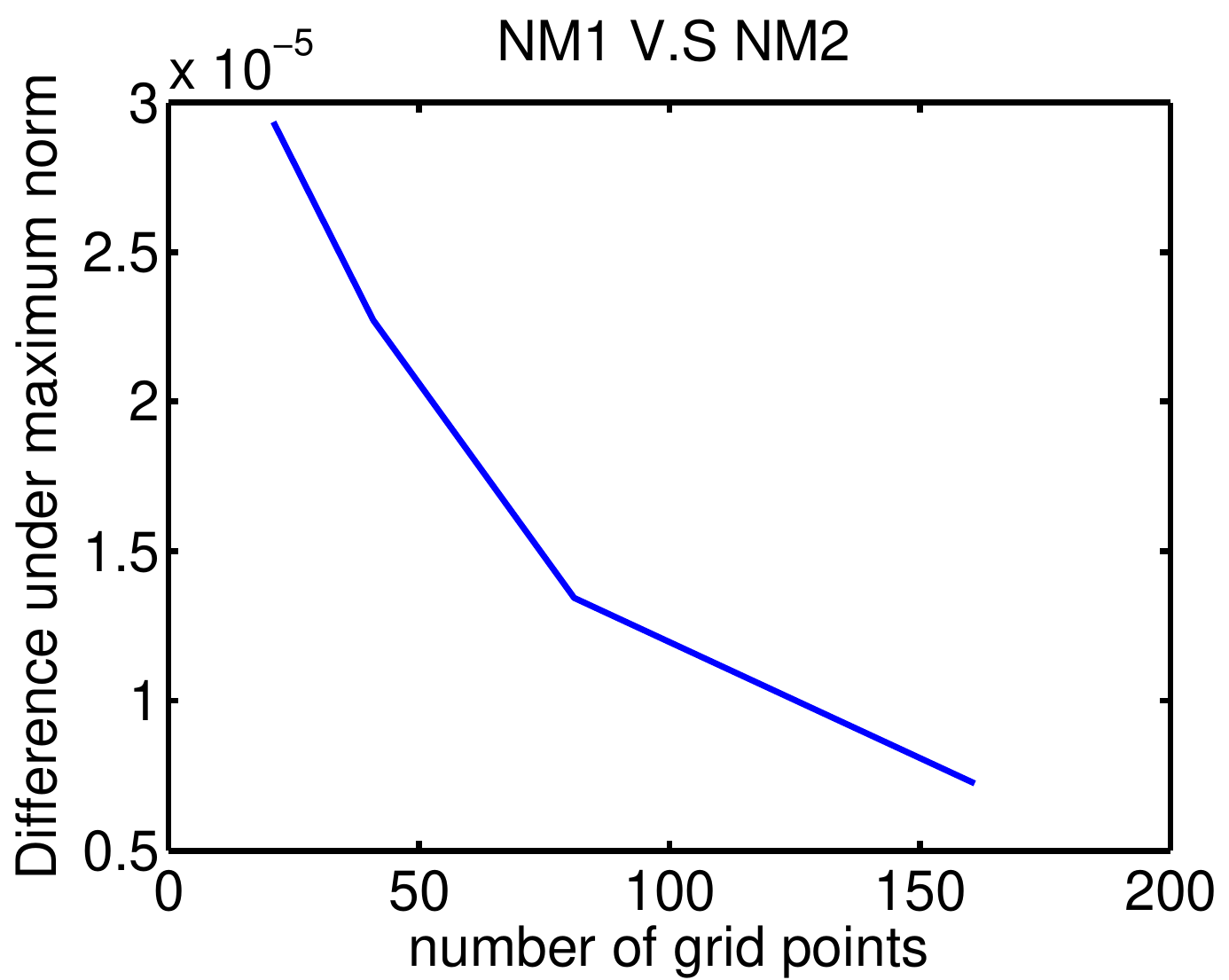}
\end{center}

\caption{Difference of the solution between NM1 and NM2 with respect to the number of gird points under the integral $l_2$  norm (left) and maximum norm (right).}
\label{norm_NM1_NM2}
\end{figure}

The main difference between algorithm NM1 and NM2 is the linearization error $O(D^2)$. Figure~\ref{norm_NM1_NM2} illustrates this error which can be reduced by refining the mesh using more grid points.  Both approaches can approximate to the same value with $\Delta S$ and $\Delta t$ small enough and can be easily applied to different nonlinear volatilities models as well as different types of options.   \\

\section{Numerical experiments}
In this section a comparison is made of two different numerical approximation methods for computing prices of European call options based on Newton's methods (NM1, NM2) and the analytic asymptotic approximation formula developed in section 2. In the asymptotic approximation formula, the Frey and Patie model (\ref{Nonlinearity}) and the RAPM model (\ref{Nonlinearity_RAPM}) are characterized by the following parameters: $(\varepsilon,\delta,\gamma)=(\rho,2,1)$ and $(\varepsilon,\delta,\gamma)=(\mu,4/3,1)$, respectively. For the finite difference Newton's methods (NM1, NM2) terminal and boundary conditions were chosen as:
\begin{displaymath}
\left\{\begin{array}{ll}
V(S,T) = (S-E)^{+}, \quad $for$ \quad 0 \leq S < S_{max},  \\
V(0,t) = 0, \quad $for$ \quad 0 \leq t \leq T,\\
V(S,t) = S - Ee^{-r(T-t)}, \quad $when$ \quad S = S_{max}.
\end{array}\right.
\end{displaymath}
Common model parameters were chosen as: $\tilde{\sigma} = 0.4, E = 100, r = 0.03, q = 0, S_{min} = 0, S_{max} = 300, T = 1/12$ and a transformation $\tau = T - t$ was used.  The tolerance for Newton's iterations was set as $tol=10^{-8}$. The initial guess in Newton's methods at the first time level was chosen as the constant value of 1. In the subsequent temporal levels the initial guess was taken from the approximate solution at the previous time level.  The fast and robust Thomas algorithm for tridiagonal solver was used in Newton's method. Calculation of integrals for the asymptotic formula was done by using the built-in Matlab function \textit{integral}.\\

\subsection{Comparison of numerical methods with explicit invariant solution}
In order to ensure all the numerical solvers mentioned in section 3 are accurate, the explicit invariant solutions for the Frey and Patie model derived by Bordag in \cite[(86), (87)]{Bordag} with parameters $c = -0.05, d_1 = 0, d_2 = 30$ were computed and taken as reference solutions for evaluating experimental order of convergence. The boundary conditions and initial conditions were generated from these invariant solutions.

The table containing the experimental order of convergence (or convergence ratio) is constructed from the convergence rate of the error defined as follows:
\[
a = \frac{\log((Err)_{m+1}/(Err)_m)}{\log((\Delta S)_{m+1}/(\Delta S)_m)}.
\]
Here the error $Err$ is defined as $Err = \Vert V(S,\tau) - \hat{V}(S,\tau) \Vert / \Vert \hat{V}(S,\tau) \Vert$ for $S \in [0.5E,1.5E]$, where $V(S,\tau)$ is the solution from numerical solver, and $\hat{V}(S,\tau)$ is from the invariant solution.  The ratio $ (\Delta S)^2 / \Delta \tau$ is fixed to be 108000, and $(\Delta S)_{m+1}/(\Delta S)_m = 0.5$.  Tables~\ref{EOC_illiquid1} and \ref{EOC_illiquid2} show results for the $l_\infty$ maximum norm and  $l_2$ integral norm.  Both results demonstrate that all the solvers converge to the same solution which converges to the explicit invariant solution with refined grid points. 

\begin{table}
\caption{EOC for the Frey and Patie Model with the $l_\infty$ maximum norm} 
\label{EOC_illiquid1}
\centering\small
\begin{tabular}{ l | l |  l  l  l  l  l  l l  }

  $\Delta \tau$ & $\Delta S$ & $Err_{NM1}$ & $a_{NM1}$ & $Err_{NM2}$ & $a_{NM2}$ & $Err_{Frozen}$ & $a_{Frozen}$ \\
  \hline\hline
  0.00833 & 30 & 2.93e-05 &   ---     &2.93e-05 &  ---    & 2.93e-05 &  \\
  0.00208 & 15 & 1.72e-06 & 4.09 & 1.72e-06 & 4.09& 1.72e-06 & 4.09\\
  5.21e-04& 7.5 & 1.02e-07 & 4.08 & 1.02e-07 & 4.08 & 1.02e-07 & 4.08 \\
  1.30e-04& 3.75 & 2.50e-08  & 2.02 & 2.50e-08  & 2.02  &2.50e-08 &2.02 \\
  3.26e-05& 1.875 & 5.00e-09  & 2.32 & 5.00e-09 & 2.32 & 5.00e-09 & 2.32 \\
  8.14e-06& 0.9375 & 1.25e-09 & 2.00 &   1.25e-09 & 2.00 & 1.25e-09 &2.00  \\
\end{tabular}
\end{table}	

\begin{table} 
\caption{EOC for the Frey and Patie Model with the $l_2$ integral norm} 
\label{EOC_illiquid2}
\centering\small
\begin{tabular}{ l | l | l  l  l  l  l  l l  }

  $\Delta \tau$ & $\Delta S$ & $Err_{NM1}$ & $a_{NM1}$ & $Err_{NM2}$ & $a_{NM2}$ &$ Err_{Frozen}$ & $a_{Frozen}$ \\
  \hline\hline
  0.00833 & 30 & 2.93e-05 &   ---     &2.93e-05 &  ---     & 2.93e-05 &  \\
  0.00208 & 15 & 1.79e-06 & 4.03 & 1.79e-06 & 4.03& 1.79e-06 & 4.03\\
  5.21e-04& 7.5 & 1.39e-07 & 3.68 & 1.39e-07 & 3.68 & 1.39e-07 & 3.68  \\
  1.30e-04& 3.75 & 4.46e-08  & 1.64 & 4.46e-08  & 1.64  &4.46e-08 &1.64 \\
  3.26e-05& 1.875 & 1.25e-08  & 1.83 & 1.25e-08 & 1.83 & 1.25e-08 & 1.83 \\
  8.14e-06& 0.9375 & 4.32e-09 & 1.53 &   4.32e-09 & 1.53 & 4.32e-09&1.53   \\
\end{tabular}
\end{table}	

\subsection{Comparison of accuracy of Newton's method and asymptotic analytic formula}
\begin{figure}
 \begin{center} 
 \includegraphics[width=0.45\textwidth]{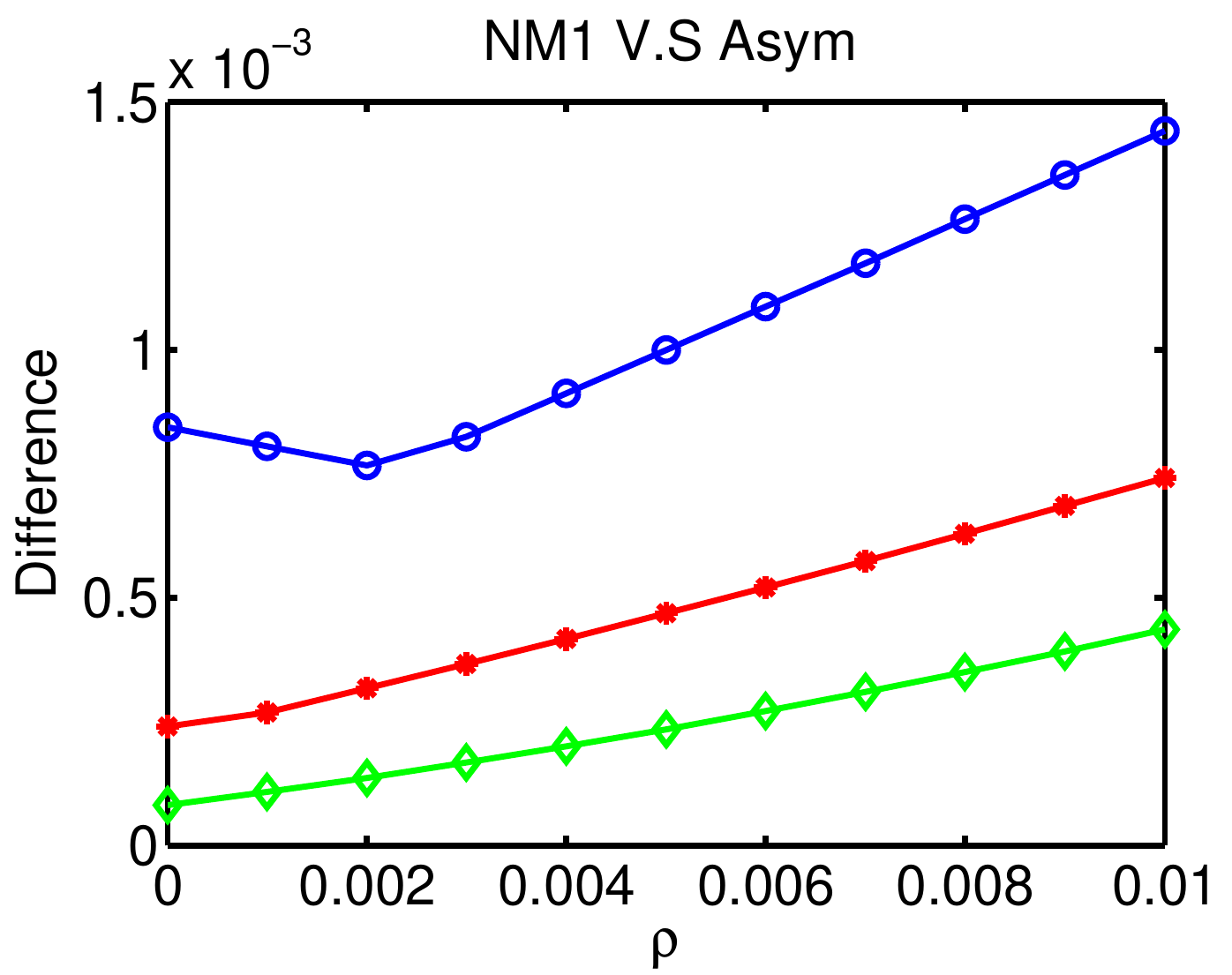}
 \includegraphics[width=0.45\textwidth]{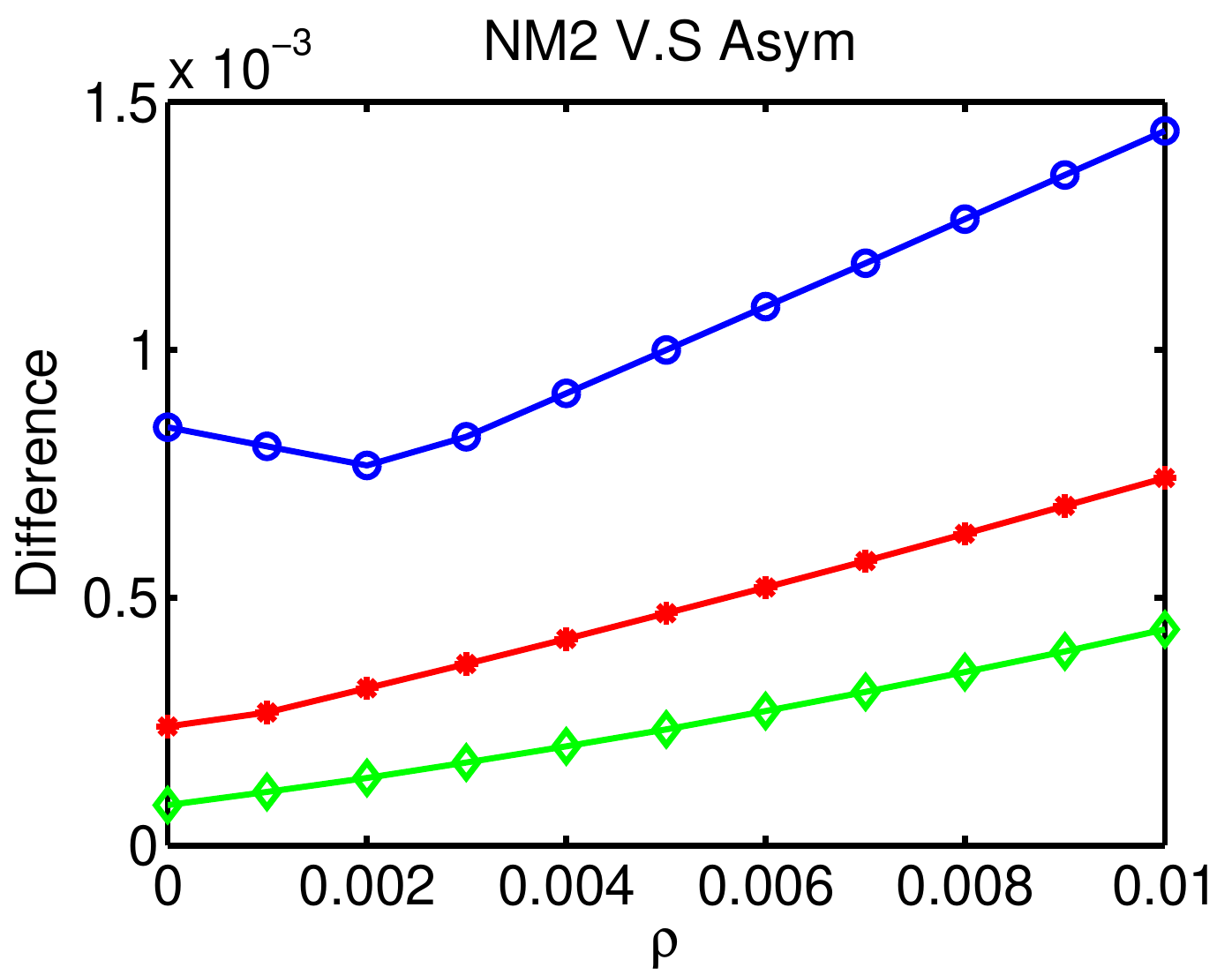}
 \\
 \includegraphics[width=0.45\textwidth]{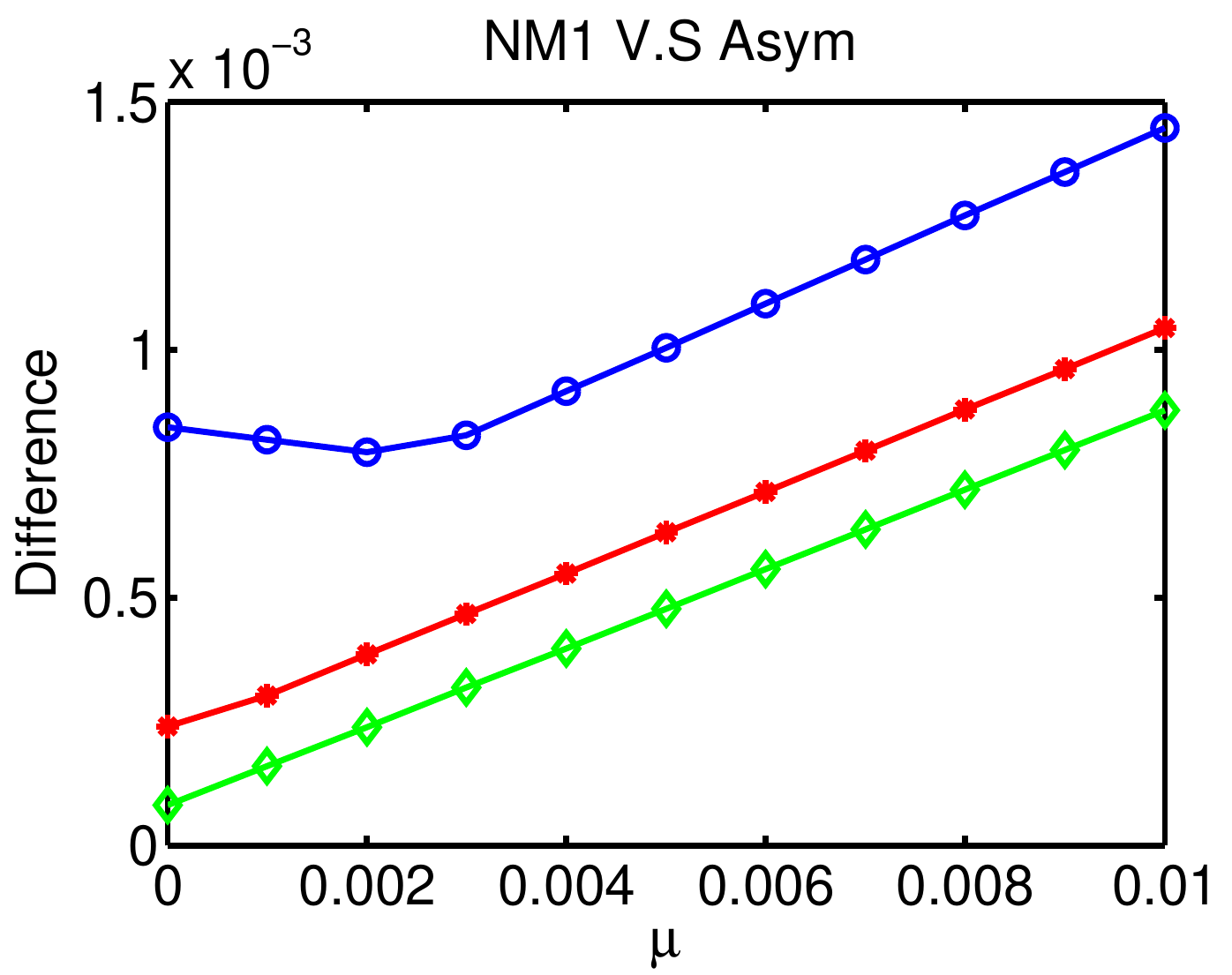}
 \includegraphics[width=0.45\textwidth]{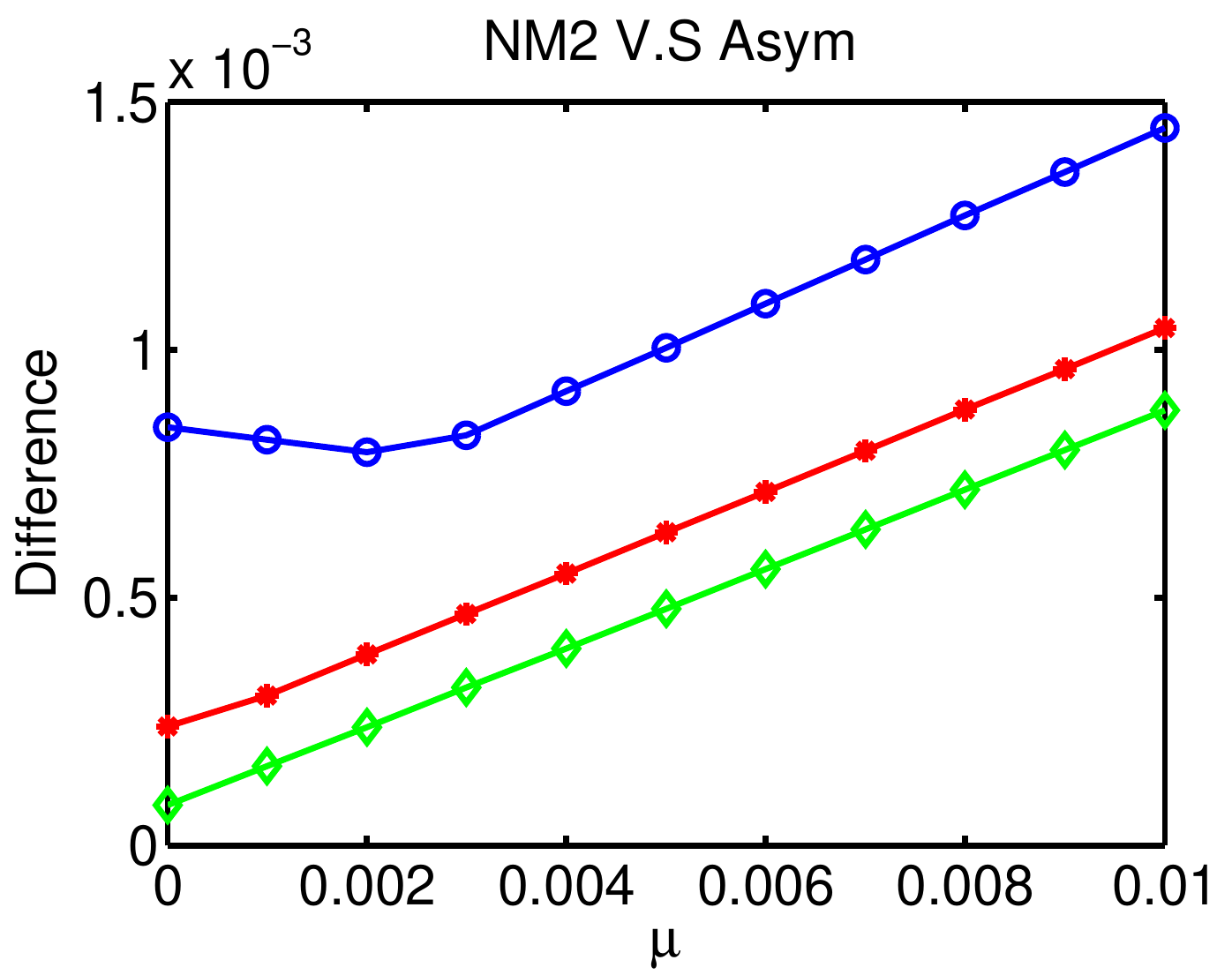}
\end{center}
 
\caption{Difference between the analytic asymptotic approximation formula and Newton's methods NM1(left), NM2(right)  for the Frey and Patie model (top row) and the RAPM model (bottom row). The circled blue line corresponds to $M=N=50$, the red line with stars corresponds to grid sizes $M=N=100$, and the green line with diamonds corresponds to $M=N=200$.}
\label{comparison_illiquid-rapm}

\end{figure}

In Figure~\ref{comparison_illiquid-rapm} errors between different methods were plotted in order to analyze the changes of the numerical approximation  with respect to different model parameter $\rho$ in the Frey and Patie model and $\mu$ in the RAPM model. The error $\Vert V(S,\tau) - \tilde{V}(S,\tau) \Vert / \Vert  \tilde{V}(S,\tau) \Vert$ for $S \in [0.5E,1.5E]$ was computed with the $l_{\infty}$ maximum norm where $V$ was calculated from Newton's method and $\tilde{V}$ was evaluated by the asymptotic formula.\\

The difference between Newton's method and the asymptotic formula can be reduced by taking smaller values of model parameters as shown in the Figure~\ref{comparison_illiquid-rapm}. When $\rho$ and $\mu$ become larger, the difference increases.  Notice that in the asymptotic formula, higher order terms such us $O(\rho^2)$ and $O(\mu^2)$ are ignored. These terms can not be neglected when considering larger values of the model parameters.

\subsection{Time complexity comparison of Newton's method and analytic asymptotic formula}
The comparison of time complexity is based on the implementation under the same Matlab computing environment in order to ensure fair comparison. Since the CPU time depends on the software implementation, the comparison is chosen to be based on calculating the so-called Experimental Order of Time Complexity $eotc$ as defined below
\[
Computation \: Time = \tilde{c} \times \Delta \tau^{eotc}
\]
and can be expressed as
\[
eotc = -\frac{\log((Time)_{n+1}/(Time)_n)}{\log((\Delta \tau)_{n+1}/(\Delta \tau)_n)}.
\]
The model parameters were chosen as $\rho = 0.005$ and $\mu = 0.005$. For all Newton based methods the spatial variable $S$ was stored in a vectorized form in order to speed up computation. The ratio of grid sizes was taken as $\Delta S/\Delta \tau = 3600$ and $(\Delta \tau)_{n+1}/(\Delta \tau)_n = 0.5$.  

Tables~\ref{table1}, \ref{table2}, \ref{table3} and \ref{table4} show the computation times and the values of EOTC. NM1,2(Fo) corresponds to computing the analytic form of the Jacobian matrix and the coefficients. NM1,2(Nu) corresponds to using a finite difference approximation of the Jacobian matrix and the coefficients. Abbreviation ,Asym' stands for results computed by means of the analytic approximation formula derived in section 2, and ,Frozen' stands for results obtained by the frozen coefficient method instead of the Newton one.\\

\begin{table}
\caption{EOTC for the Frey and Patie model} 
\label{table1}
\centering\small
\begin{tabular}{ l | l | l  l | l  l | l  l  }
  $\Delta \tau$ & $\Delta S$ & NM1(Fo)& $eotc$ & NM2(Fo) & $eotc$ & Asym & $eotc$ \\
                &            &  (sec) &        &   (sec) &        & (sec)&        \\
  \hline
  0.00208 & 7.5 & 0.053 &   ---     & 0.041 &  ---     & 0.291 & \\
  0.00104 & 3.75 & 0.121 & 1.190 & 0.101 & 1.300 & 0.467 & 0.682\\
  5.21e-04& 1.875 & 0.524 & 2.114 & 0.292 & 1.531 & 0.826 & 0.822 \\
  2.60e-04& 0.9375 & 4.748 & 3.179 & 1.544 & 2.402 & 1.845 & 1.159\\
  1.30e-04& 0.4687 & 70.32 & 3.888 & 17.06 & 3.465 & 4.549 & 1.301 \\
\end{tabular}
\end{table}	

\begin{table}
\caption{EOTC for the Frey and Patie model} 
\label{table2}
\centering\small
\begin{tabular}{ l | l | l  l | l  l | l  l  }

  $\Delta \tau$ & $\Delta S$ & NM1(Nu) & $eotc$ & NM2(Nu) & $eotc$ & Frozen & $eotc$ \\
                &            &  (sec)  &        &   (sec) &        & (sec)  &        \\
  \hline
  0.00208 & 7.5 & 0.170 &   ---     & 0.220 &  ---     & 0.040 & \\
  0.00104 & 3.75 & 0.541 & 1.670 & 0.503 & 1.193 & 0.095 & 1.247\\
  5.21e-04& 1.875 & 4.308 & 2.993 & 1.791 & 1.832 & 0.301 & 1.663 \\
  2.60e-04& 0.9375 & 25.84 & 2.584 & 11.09 & 2.630 & 1.653 & 2.457\\
  1.30e-04& 0.4687 & 230.0 & 3.153 & 95.69 & 3.108 & 17.91 & 3.437 \\
  \hline
\end{tabular}
\end{table}

\begin{table} 
\caption{EOTC for the RAPM Model} 
\label{table3}
\centering\small
\begin{tabular}{ l | l | l  l | l  l | l  l  }

  $\Delta \tau$ & $\Delta S$ & NM1(Fo) & $eotc$ & NM2(Fo) & $eotc$ & Asym & $eotc$ \\
                &            &  (sec)  &        &   (sec) &        & (sec)  &        \\
  \hline
  0.00208 & 7.5 & 0.060 &   ---     & 0.133 &  ---     & 0.353 & \\
  0.00104 & 3.75 &  0.157 & 1.387 & 0.613 & 2.204 & 0.580 & 0.716\\
  5.21e-04& 1.875 & 0.585 & 1.897 & 3.360 & 2.454 & 1.104 & 0.928 \\
  2.60e-04& 0.9375 & 4.918 & 3.071 & 27.31 & 3.023 & 2.488 & 1.172\\
  1.30e-04& 0.4687 & 66.76 & 3.762 & 224.5 & 3.039 & 6.171 & 1.310 \\
  \hline
\end{tabular}
\end{table}	

\begin{table}
\caption{EOTC for the RAPM Model} 
\label{table4}
\centering\small
\begin{tabular}{ l | l | l  l  l  l  l  l  }

  $\Delta \tau$ & $\Delta S$ & NM1(Nu) & $eotc$ & NM2(Nu) & $eotc$ & Frozen & $eotc$ \\
                &            &  (sec)  &        &   (sec) &        & (sec)  &        \\
  \hline
  0.00208 & 7.5 & 0.457 &  ---      & 0.426 & ---      & 0.032 & \\
  0.00104 & 3.75 & 1.612 & 1.818 & 1.717 & 2.010 & 0.090 & 1.491\\
  5.21e-04& 1.875 & 10.56 & 2.711 & 9.280 & 2.434 & 0.306 & 1.765 \\
  2.60e-04& 0.9375 & 58.95 & 2.480 & 70.59 & 2.927 & 1.735 & 2.503\\
  1.30e-04& 0.4687 & 465.2 & 2.980 & 588.4 & 3.059 & 16.95 & 3.288 \\
  \hline
\end{tabular}
\end{table}

\begin{figure}
\begin{center}
 \includegraphics[width=0.45\textwidth]{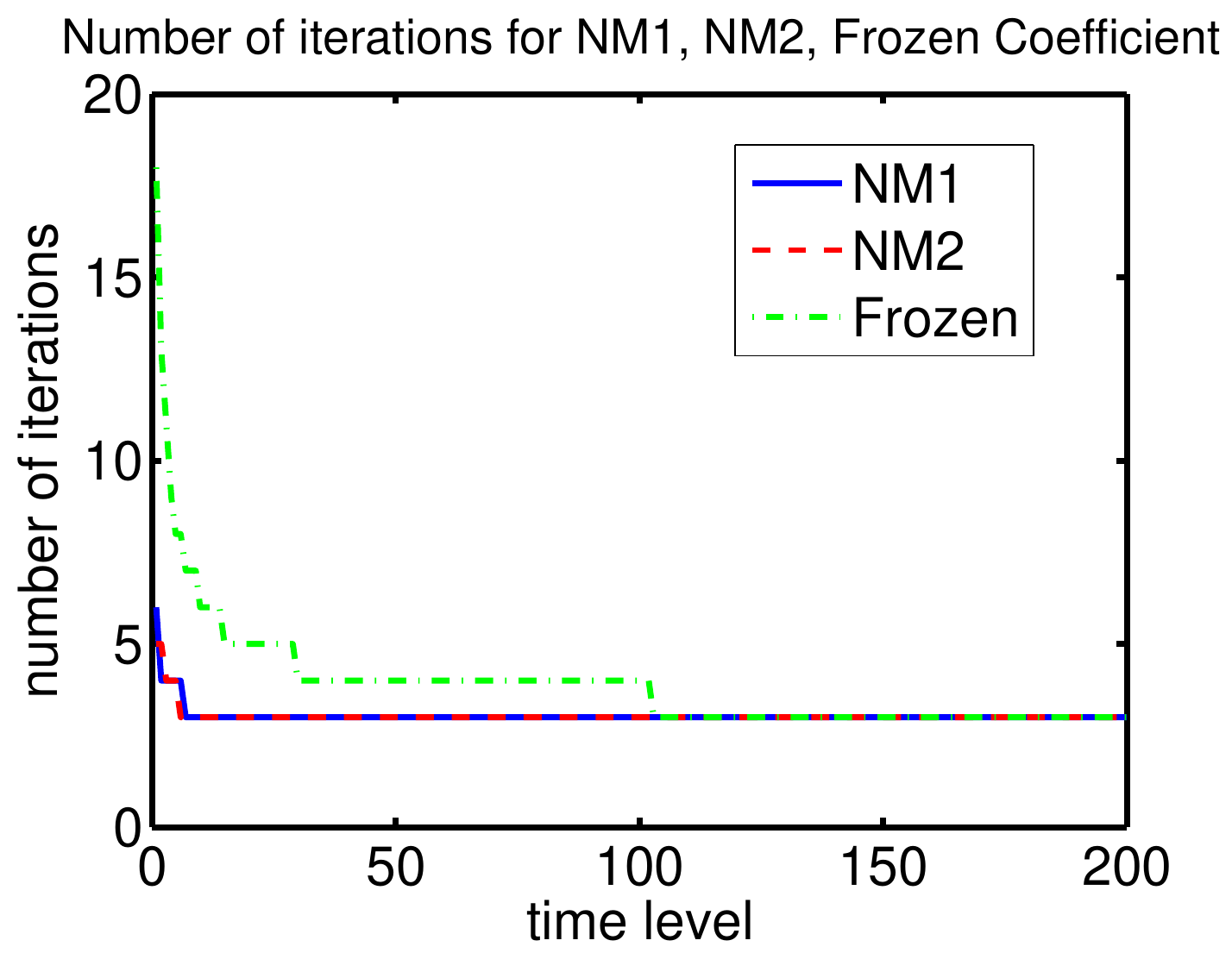}
\end{center}
 
\caption{Number of iterations (vertical axis) for using NM1 (Blue), NM2 (Green) and frozen coefficient (Red) in the Frey and Patie model for different times (horizontal axis). }
\label{N_ite}
\end{figure}

The results from evaluating the computational complexity and the experimental order of time complexity shows that the analytic approximation formula has the advantage when considering smaller time steps $\Delta \tau$. Hence it can be successfully adopted for model calibration using high frequency data.  When all the numerical methods converge, Newton's method seems to have worse performance when compared to frozen coefficients methods as can be seen from Figure~\ref{N_ite} which shows the number of iterates for the example with grid points $M=N=200$ and $\rho=0.01$. Clearly, for the first few time levels, the method of frozen coefficients requires higher number of iterates to ensure convergence.  However, since the solution from previous time level is taken to be the initial guess for the new time level it helps to reduce the number of iterates for the subsequent time levels. Newton's based methods spent most of the time by evaluating Jacobian matrices.  A possible improvement is to combine Newton's method and frozen coefficients method, or by implementing Broyden's type of updates for the Jacobian matrix.
 
\section{Calibration of the Frey and Patie model to market quotes data}
\begin{figure}
\begin{center}
\includegraphics[width=0.45\textwidth]{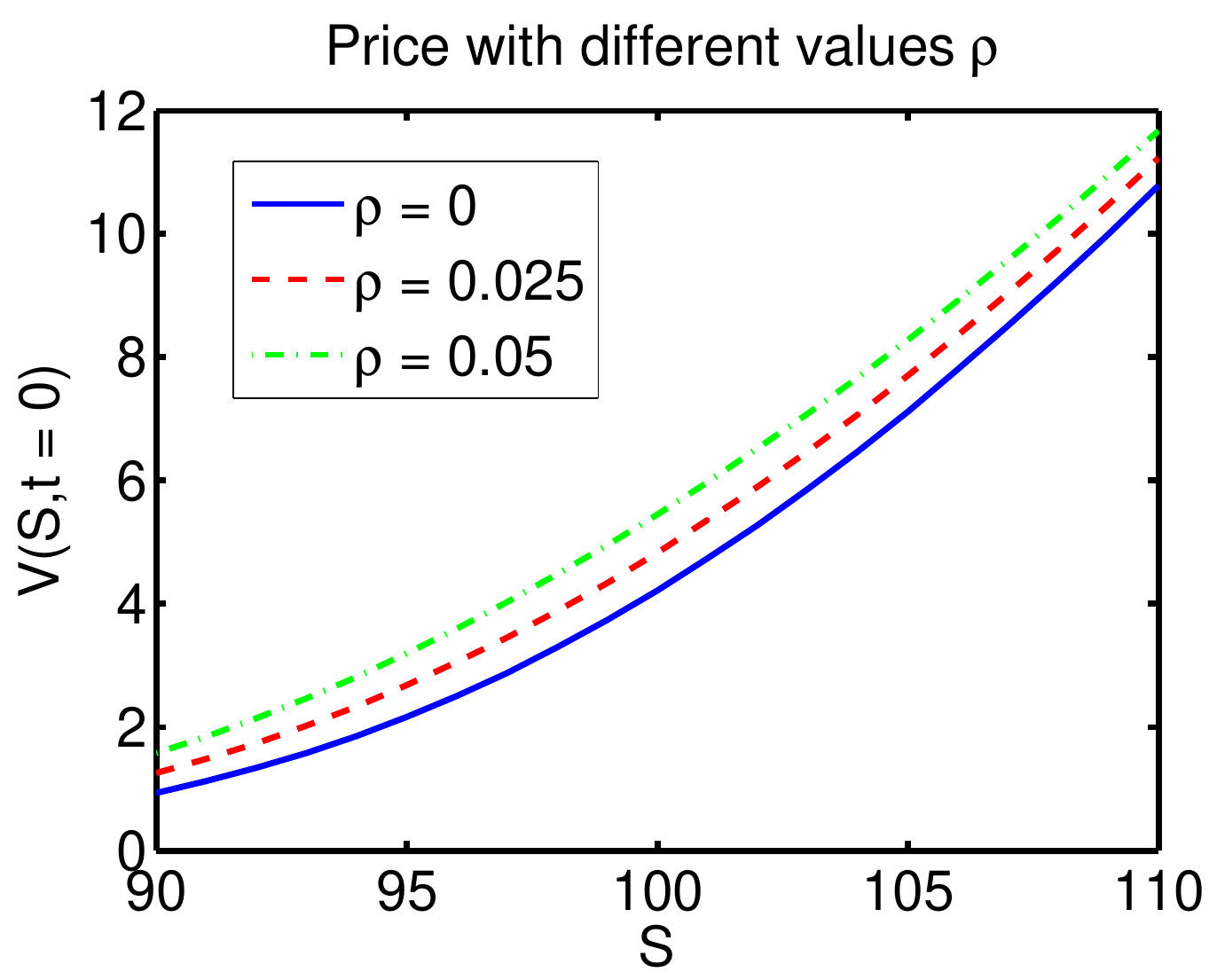}
\includegraphics[width=0.45\textwidth]{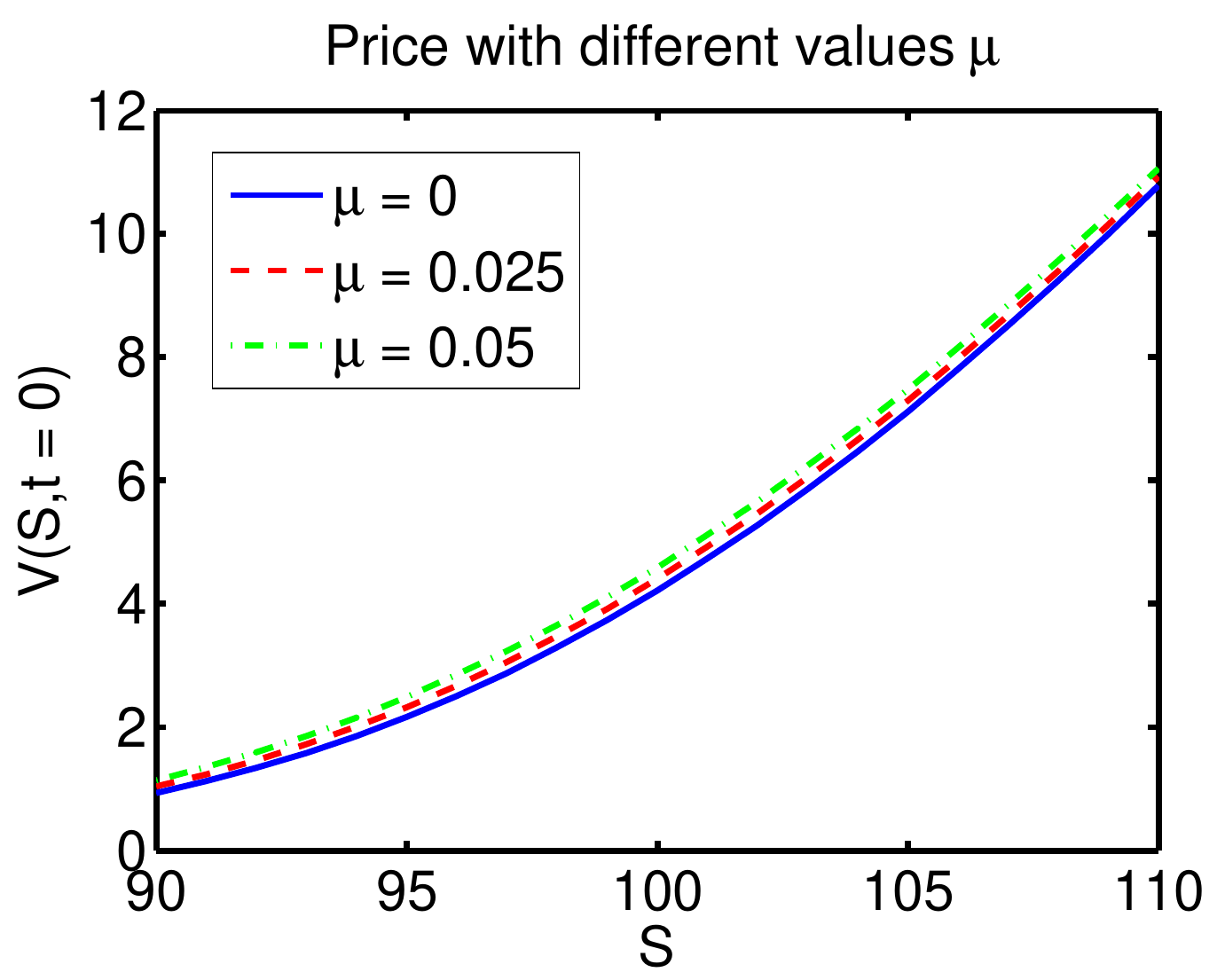}
\end{center}
\caption{Different option prices for $\rho = 0, 0.025, 0.05$ in the Frey and Patie model (left) and for $\mu = 0, 0.025, 0.05$ in the RAPM model (right).}
\label{Diff_Prices}
\end{figure}

Numerical results from section 4 have demonstrated that the asymptotic formula can be used for accurate approximation of a solution to the nonlinear Black-Scholes equation if the parameters $\rho$ and $\mu$ are sufficiently small.  From Figure \ref{Diff_Prices}, it is important to notice that the option price increases for asset prices close to $E$ when these parameters are increasing. In fact $\rho$ and $\mu$ can be calibrated using market data to observe whether the market of underlying asset has high or low liquidity.
\\

In the calibration experiments the parameter $\rho$ for the Frey and Patie model was calibrated by using the call option time series from Apple Inc. (AAPL) in NASDAQ quotes market. Bisection method was used in the search algorithm as described in Algorithm~\ref{Model_Calibration}.  The parameters in the calibration process were fixed as $r = 0.01$, $E = 106$, $q = 0$, and $\tilde\sigma=\sigma_{impl}$ was computed as the implied volatility from the market quotes prices. As for the solution method (,Solver') both analytic approximation formula and Newton based methods were used.  Table~\ref{Cali} shows similar calibration results for both methods when the parameter $\rho$ is not large using these market data.  This means that the analytic approximation formula has a benefit of performing fast calibration when compared to Newton's method.

\begin{center}
 \includegraphics{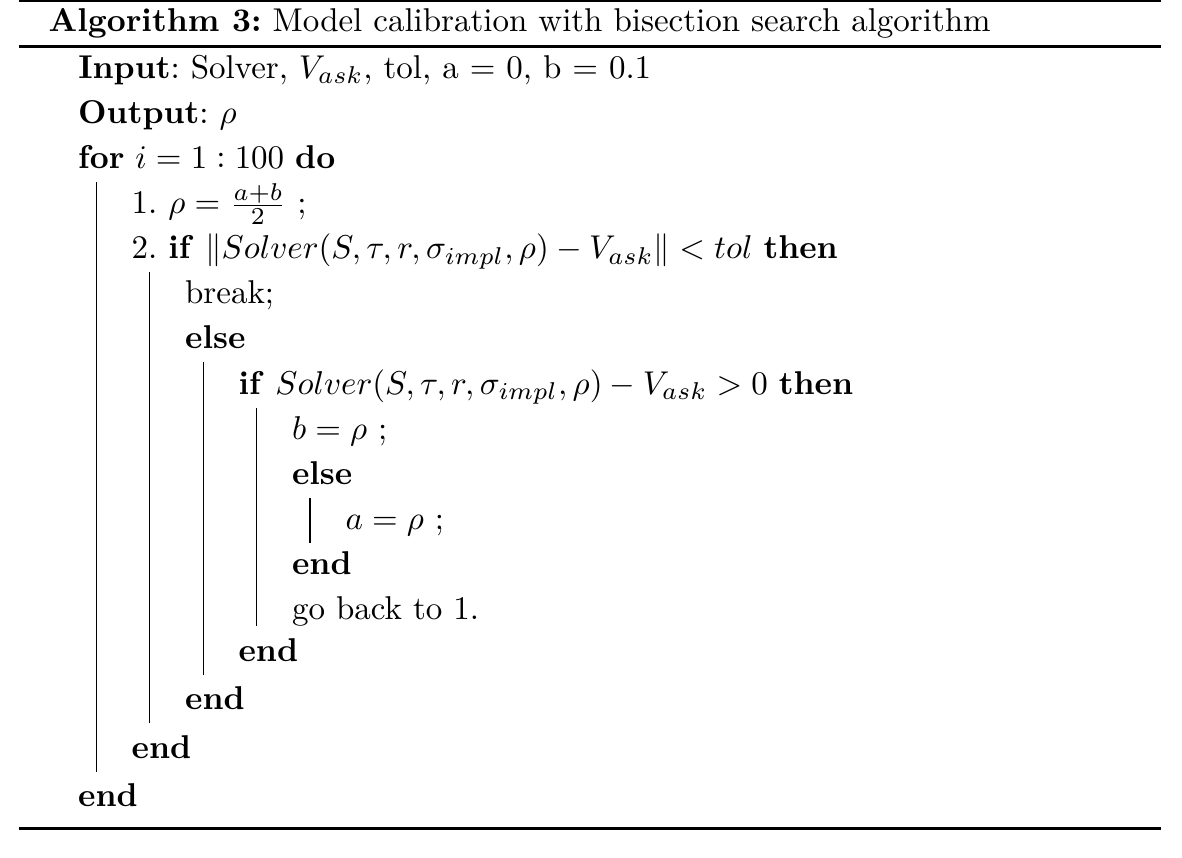}
\end{center}

\begin{table} 
\caption{Calibration results} 
\label{Cali}
\centering\small
\begin{tabular}{ l | l  l  l  l  l  l  l  }

  $\tau$ & $S$ &  $V_{bid}$ & $V_{ask}$ &  $\sigma_{impl}$ & $\rho_{Asym}$& $\rho_{Newton}$ \\
  \hline\hline
  0.0753 & 107.67 & 6.100 & 6.200 & 0.443 & 3.807e-03 & 3.956e-03 \\
  0.0674 & 107.14 & 4.925 & 5.000 & 0.389 & 2.848e-03 & 2.934e-03 \\
  0.0595 & 112.37 & 8.225 & 8.300 & 0.401 & 3.492e-03 & 3.584e-03 \\
  0.0515 & 111.70 & 7.625 & 7.700 & 0.419 & 3.383e-03 & 3.347e-03 \\
  0.0436 & 109.01 & 6.225 & 6.300 & 0.506 & 2.939e-03 & 3.030e-03 \\
  0.0357 & 107.58 & 4.525 & 4.600 & 0.455 & 2.875e-03 & 2.995e-03 \\
  0.0277 & 110.37 & 5.950 & 6.000 & 0.458 & 2.228e-03 & 2.247e-03 \\
  0.0198 & 113.28 & 8.300 & 8.350 & 0.569 & 2.847e-03 & 2.912e-03 \\
\end{tabular}
\end{table}

\section{Conclusion}
In this paper two different linearization numerical methods for solving the nonlinear Black-Scholes equation are proposed and analyzed. Numerical results are compared in their accuracy and time complexity for the Frey and Patie illiquid market model and the risk-adjusted pricing methodology model. It turns out that the analytic approximation formula is more suitable for computation when the model parameters are sufficiently small. In particular it can be applied in calibrating parameters using market data efficiently as it is a time consuming process for a full temporal-spatial finite difference approximation scheme based on Newton's method. On the other hand, the analytic approximation formula becomes restrictive as the error increases when the parameters become larger. The Newton's method is easy to implement and suits various types of nonlinear Black-Scholes equations.  There are different approaches to implement Newton's method and two of them are discussed in this paper.  Although time complexity is a general problem, it can be improved by combining other techniques or by using the so-called Newton-like methods to approximate the Jacobian matrix in order to reduce the number of iterates.  Both techniques in fact can be extended to solve other types of nonlinear option pricing models, and the resulting numerical solutions may also be considered as a benchmark solution when exact solutions do not exist.

\section*{Acknowledgements}
 
The authors thank Prof. Ljudmila A. Bordag for her kind suggestion about using the invariant solutions.

This research is supported by the European Union in the FP7-PEOPLE-2012-ITN project STRIKE - Novel Methods in Computational Finance (304617) and the Slovak research Agency Project VEGA 1/0780/15.

\section{Appendix, proof of Theorem~\ref{maintheorem}}

A solution $u(x,\tau)$ to the non-homogeneous parabolic PDE
\begin{equation*}\label{vseobuloha}
   \left\{
     \begin{array}{ll}
       \frac{\partial u}{\partial \tau}-a^2\frac{\partial^2 u}{\partial x^2}=f(x,\tau),
       &(x,\tau)\in\mathds{R}\times(0,\infty)\\
       u(x,0)=0,
       &x\in\mathds{R}
     \end{array}
   \right.
\end{equation*}
is given by the variation of constant formula and is given by
\[
u(x,\tau)=\int_0^\tau\int_{-\infty}^\infty G(x-\xi,\tau-s)f(\xi,s)d\xi\, ds, \quad \hbox{where}\ 
G(x,\tau)=\frac{1}{\sqrt{4\pi a^2\tau}}e^{-\frac{x^2}{4a^2\tau}}.
\]

The solution of equation (\ref{zakluloha}) can be written
\begin{align} 
  \begin{split}\nonumber
    u(x,\tau)=\int_0^\tau\int_{-\infty}^\infty &\frac{1}{\sqrt{2\pi\tilde{\sigma}^2(\tau-\xi)}} e^{-\frac{(x-s)^2}{2\tilde{\sigma}^2(\tau-\xi)}}\frac{E^\gamma}{(2\pi\tilde{\sigma}^2\xi)^{\frac{\delta}{2}}}A(\xi)
    \\ 
    &\times e^{-\frac{\delta}{2\tilde{\sigma}^2\xi}s^2+\left[\gamma-\delta-\alpha(1-\delta)\right]s -\left[\beta+q\delta+\frac{\delta}{2}(1-\alpha)^2\tilde{\sigma}^2\right]\xi}ds\,d\xi.
  \end{split}
\end{align}
Let us consider the change of variables in (\ref{simplify}) and introduce the function:
\begin{equation}\label{simplifyR}
    R(\xi)=\left[\beta+q\delta+\frac{\delta}{2}(1-\alpha)^2\tilde{\sigma}^2\right]\xi = -\beta(\delta-1)\xi
\end{equation}
because $\beta = -\frac{\tilde{\sigma}^2}{2}\alpha^2 - r$ (see (\ref{alfa_beta})). 

In order to simplify further notation and let $EXP$ denote the power of the exponential function, i.e.		
\begin{align}
  \begin{split}\nonumber
    EXP=&-\frac{x^2-2xs+s^2}{2\tilde{\sigma}^2(\tau-\xi)}-\frac{\delta}{2\tilde{\sigma}^2\xi}s^2+Ps-R(\xi)\\
    =&-\frac{\xi+\delta(\tau-\xi)}{2\tilde{\sigma}^2(\tau-\xi)\xi}s^2+\left[\frac{x}{\tilde{\sigma}^2(\tau-\xi)}+P\right]s-R(\xi) -\frac{x^2}{2\tilde{\sigma}^2(\tau-\xi)}\\
    =&-\frac{Q(\tau,\xi)}{2\tilde{\sigma}^2(\tau-\xi)\xi}\left\{s^2-2\frac{x+P\tilde{\sigma}^2(\tau-\xi)}{Q(\tau,\xi)}\xi s+\left[\frac{x+P\tilde{\sigma}^2(\tau-\xi)}{Q(\tau,\xi)}\xi\right]^2\right.\\
    &-\left.\left[\frac{x+P\tilde{\sigma}^2(\tau-\xi)}{Q(\tau,\xi)}\xi\right]^2\right\}-R(\xi)-\frac{x^2}{2\tilde{\sigma}^2(\tau-\xi)}\\
    =&-\frac{Q(\tau,\xi)}{2\tilde{\sigma}^2(\tau-\xi)\xi}\left\{s-\frac{x+P\tilde{\sigma}^2(\tau-\xi)}{Q(\tau,\xi)}\xi\right\}^2+ \frac{\left[x+P\tilde{\sigma}^2(\tau-\xi)\right]^2}{2\tilde{\sigma}^2(\tau-\xi)Q(\tau,\xi)}\xi\\
    &-R(\xi) -\frac{x^2}{2\tilde{\sigma}^2(\tau-\xi)}.
  \end{split}
\end{align}
Consider the  function  $\Lambda(\tau,\xi)$ defined in (\ref{simplify}). Then the inner integral can be calculated as follows:
\begin{align*}
  \begin{split}\nonumber
    u(x,\tau)=\int_0^\tau & \frac{E^\gamma A(\xi)}{\Lambda(\tau,\xi)} \exp\left\{\frac{\left[x+ P\tilde{\sigma}^2(\tau-\xi)\right]^2} {2\tilde{\sigma}^2(\tau-\xi)Q(\tau,\xi)}\xi-R(\xi)-\frac{x^2}{2\tilde{\sigma}^2(\tau-\xi)}\right\} 
    \\ 
  & \hspace{-0.50cm} \times \int_{-\infty}^\infty\frac{1}{\sqrt{\frac{2\pi\tilde{\sigma}^2(\tau-\xi)\xi}{Q(\tau,\xi)}}} \exp\left\{-\frac{Q(\tau,\xi)}{2\tilde{\sigma}^2(\tau-\xi)\xi}\left[s-\frac{x+P\tilde{\sigma}^2(\tau-\xi)} {Q(\tau,\xi)}\xi\right]^2\right\}ds\,d\xi \\
    =\int_0^\tau &\frac{E^\gamma A(\xi)}{\Lambda(\tau,\xi)} \exp\left[-\frac{\delta x^2}{2\tilde{\sigma}^2Q(\tau,\xi)}+\frac{P x\xi}{Q(\tau,\xi)}+ \frac{P^2\tilde{\sigma}^2(\tau-\xi)\xi} {2Q(\tau,\xi)}-R(\xi)\right] d\xi.
  \end{split}
\end{align*}
Hence
\begin{align}
  \begin{split}\nonumber
    u(x,\tau)=\int_0^\tau &\frac{E^\gamma A(\xi)}{\Lambda(\tau,\xi)} \exp\left\{\frac{\xi-\left[\delta\tau+(1-\delta)\xi\right]}{2\tilde{\sigma}^2(\tau-\xi) \left[\delta\tau+(1-\delta)\xi\right]}x^2+\frac{\left[\gamma-\delta-\alpha(1-\delta)\right] x \xi}{\delta\tau+(1-\delta)\xi}\right.\\
    &\left. +\frac{\left[\gamma-\delta-\alpha(1-\delta)\right]^2\tilde{\sigma}^2(\tau-\xi)\xi} {2\left[\delta\tau+(1-\delta)\xi \right]}+\beta(\delta-1)\xi\right\}d\xi.
  \end{split}
\end{align}

Now let us consider the case $\delta\neq1$. Since
\begin{align}
  \begin{split}\nonumber
    \frac{\xi}{\delta\tau+(1-\delta)\xi}&=\frac{1}{1-\delta} \frac{\delta\tau+(1-\delta)\xi-\delta\tau}{\delta\tau+(1-\delta)\xi}=\frac{1}{1-\delta} -\frac{\delta\tau}{1-\delta}\frac{1}{\delta\tau+(1-\delta)\xi}, \\
    \frac{(\tau-\xi)\xi}{\delta\tau+(1-\delta)\xi}&=B\xi+C+\frac{D}{\delta\tau+(1-\delta)\xi},
  \end{split}
\end{align}
where $B=\frac{1}{1-\delta}, C=\frac{\tau}{(1-\delta)^2}$ and $D=-\frac{\delta\tau^2}{(1-\delta)^2}$. Therefore
\begin{align}
  \begin{split}\nonumber
    u(x,\tau)=\int_0^\tau &\frac{E^\gamma A(\xi)}{\Lambda(\tau,\xi)} \exp\left\{-\frac{\delta x^2}{2\tilde{\sigma}^2}\frac{1}{Q(\tau,\xi)}+\frac{P x}{1-\delta}-\frac{Px\delta\tau}{(1-\delta)Q(\tau,\xi)}+ \frac{P^2\tilde{\sigma}^2\xi} {2(\delta-1)}\right.\\
    &\left. + \frac{P^2\tilde{\sigma}^2\tau}{2(1-\delta)^2}- \frac{P^2\tilde{\sigma}^2\delta\tau^2}{2(1-\delta)^2 Q(\tau,\xi)}+\beta(\delta-1)\xi\right\}d\xi\\
    =\int_0^\tau &\frac{E^\gamma A(\xi)}{\Lambda(\tau,\xi)}\exp\left\{\left[\frac{P^2\tilde{\sigma}^2}{2(\delta-1)}+\beta(\delta-1)\right]\xi\right.\\
    &\left. + \frac{Px}{1-\delta}+ \frac{P^2\tilde{\sigma}^2\tau}{2(1-\delta)^2}-\left[\frac{\delta x^2}{2\tilde{\sigma}^2}+ \frac{Px\delta\tau}{1-\delta}+ \frac{P^2\tilde{\sigma}^2\delta\tau^2}{2(1-\delta)^2}\right] \frac{1}{Q(\tau,\xi)}\right\}d\xi.
  \end{split}
\end{align}
Substituting the terms $P,Q(\tau,\xi),\Lambda(\tau,\xi)$ yields the form of the solution $u(x,\tau)$ as stated in Theorem 2.1. 


\begin{thebibliography}{30}
\bibitem{AAMR} P. Amster, C.G. Averbuj, M.C. Mariani, D. Rial, A Black--Scholes option pricing model with transaction costs. \emph{J. Math. Anal. Appl.},
{303} (2005),
688-695.

\bibitem{doplnkyAP} M. Avellaneda, A. Par\'as, \newblock Dynamic hedging portfolios for derivative securities in the presence of large transaction costs. \newblock {\em Appl. Math. Finance}, 
1 (1994), 
165-193.

\bibitem{BaSo} G. Barles, M.-H. Soner, \newblock Option pricing with transaction costs and a nonlinear {B}lack-{S}choles equation. \newblock {\em Finance Stoch.}, 
2 (1998), 
369-397.

\bibitem{BS} F. Black, M.S. Scholes, The Pricing of options and corporate liabilities, \textit{J. Political Economy},
81 (1973), 
637-654.

\bibitem{Bordag} L.A. Bordag, R. Frey,  Pricing options in illiquid markets: symmetry reductions and exact solutions, \emph{ Nonlinear Models in Mathematical Finance: New Research Trends in Option Pricing}. Nova Science Publishers, Inc. New York (2008), pp.103-130.

\bibitem{BordagBook} L.A. Bordag, Geometrical Properties of Differential Equations: Applications of the Lie Group Analysis in Financial Mathematics, World Scientific, (2015).

\bibitem{Rafael} R. Company, L. J\'{o}dar, J.R. Pintos, A consistent stable numerical scheme for a nonlinear option pricing model in illiquid markets, \textit{Math. Comput. Simul.},
82 (2012),
1972-1985.

\bibitem{CompanyNavaro} R. Company, E. Navarro, J.R. Pintos, and F. Ponsoda, Numerical solution of linear and nonlinear Black-Scholes option pricing equations. \textit{Comput. Math. Appl.}, 
{56} (2008),
813-821.
\bibitem{Matthias} M. Ehrhardt, Nolinear Models in Mathematical Finance, Nova Science Publishers, Inc. New York (2008).

\bibitem{Matthias2} M. Ehrhardt, R. Valkov, Numerical analysis of nonlinear European option pricing problem in illiquid markets, 
Preprint BUW-IMACM 14/23 (2014).

\bibitem{Frey1998} R. Frey, \newblock Perfect option hedging for a large trader, \newblock {\em Finance Stoch.}, 
2 (1998), 
115-142.

\bibitem{Frey} R. Frey, Market Illiquidity as a Source of Model Risk in Dynamic Hedging in Model Risk, RISK Publication, E. Gilbson Ed., London (2000).

\bibitem{FP} R. Frey, P. Patie, \newblock Risk management for derivatives in illiquid markets: a simulation study, \newblock In {\em Advances in finance and stochastics}. Springer, Berlin, (2002) 137-159.

\bibitem{FS} R. Frey, A. Stremme, \newblock Market volatility and feedback effects from dynamic hedging. \newblock {\em Math. Finance}, 
7 (1997), 
351-374.

\bibitem{Heider} P. Heider, Numerical methods for nonlinear Black-Scholes equations, \textit{Appl. Math. Finance}, 
17 (2010),
59-81.

\bibitem{HWW} T. Hoggard, A.E. Whalley, P. Wilmott, \newblock Hedging option portfolios in the presence of transaction costs, \newblock \emph{ Advances in Futures and Options Research},
7 (1994), 
21-35.

\bibitem{holmes2} M. H. Holmes, Introduction to Perturbation Methods, Texts in Applied Mathematics vol. 20, Springer (2013).

\bibitem{H} J.C. Hull, \newblock {\em Options, Futures, and Other Derivatives (5th Edition)}, \newblock {Prentice Hall}, 2002.

\bibitem{Daniel2} M. Janda\v{c}ka, D. \v{S}ev\v{c}ovi\v{c}, On the risk-adjusted pricing-methodology-based valuation of vanilla options and explanation of the volatility smile, \textit{J. Appl Math}
3 (2005),
235-258.

\bibitem{Koleva} M.N. Koleva, L. G. Vulkov, Quasilinearization numerical scheme for fully nonlinear parabolic problems with applications in models of mathematical finance, \textit{Math. Comput. Modelling}, 
57 (2013),
2564--2575.

\bibitem{doplnkyKr} M. Kratka, \newblock No mystery behind the smile, \newblock {\em Risk },
9 (1998), 
67-71.

\bibitem{Kw} Y.K. Kwok, \newblock {\em Mathematical models of financial derivatives}, \newblock Springer Finance. Springer-Verlag Singapore, Singapore, 1998.

\bibitem{doplnkyLe} H.E. Leland, \newblock Option pricing and replication with transaction costs, \newblock {\em J. Finance},
40 (1985), 
1283-1301.

\bibitem{SW} P.J. Sch{\"o}nbucher, P. Wilmott, \newblock The feedback effect of hedging in illiquid markets, \newblock {\em SIAM J. Appl. Math.}, 
61 (2000) 232-272.

\bibitem{Daniel1} D. \v{S}ev\v{c}ovi\v{c}, B. Stehl\'ikov\'a, K. Mikula, Analytical and Numerical Methods for Pricing Financial Derivatives, Nova Science Publishers, New York, (2011).


\bibitem{SZ}  D. \v{S}ev\v{c}ovi\v{c}, M. \v{Z}it\v{n}ansk\'a, Analysis of the nonlinear option pricing model under variable transaction costs, submitted, (2015).


\end{thebibliography}
\end{document}